\documentclass[twocolumn]{aastex63}

\usepackage{graphicx,textcomp,fancyhdr,hyperref,xcolor,fontawesome}
\definecolor{linkcolor}{rgb}{0.1216,0.4667,0.7059}
\usepackage[caption=false]{subfig}
\usepackage{float}
\usepackage{footnote}

\shortauthors{Michelle L. Hill et al.}

\begin{document}

\title{Smaller Than Earth Habitability Model (STEHM): The Lower Size Limit for Atmosphere Retention in the Habitable Zone} 

\author[0000-0002-0139-4756]{Michelle L. Hill}
\affiliation{Department of Earth and Planetary Sciences, University of California Riverside, Riverside, CA 92521, USA}
\affiliation{Department of Earth and Planetary Sciences, Stanford University, Stanford, CA 94305, USA}
\email{MichelleLHill@stanford.edu}
  
\author[0000-0002-7084-0529]{Stephen R. Kane}
\affiliation{Department of Earth and Planetary Sciences, University of California Riverside, Riverside, CA 92521, USA}

\author[0000-0002-6943-3192]{Bradford J. Foley}
\affiliation{Department of Geosciences, Pennsylvania State University,
  University Park, PA 16802, USA}

\author[0000-0003-2915-5025]{Laura K. Schaefer}
\affiliation{Department of Earth and Planetary Sciences, Stanford University, Stanford, CA 94305, USA}

%%%%%%%%%%%%%%%%%%%%%%%%%%%%%%%%%%%%%%%%%%%%%%%%%%%%%%%%%%%%%%%%%%%%

\begin{abstract}

With recent advances in exoplanet observational techniques enabling the discovery of increasingly smaller planets, a crucial question emerges in the search for habitable planets: how small can a planet be and still maintain an atmosphere?
We present results from the Smaller Than Earth Habitability Model (STEHM) which examines how small a planet can be and still maintain a long-term (multi-gigayear) atmosphere for planets from 1.0~$R_\oplus$ down to 0.5~$R_\oplus$. The model is based on a stagnant lid planet orbiting within the habitable zone of a sun-like star.
Our model demonstrates that planets $\geq$0.8~$R_\oplus$ can maintain their atmospheres under our Earth-like default conditions for a solar analog star, while smaller planets lose their atmospheres. Variations from the default Earth-like values cause mostly minor variations to the planet size boundary results, with some changes allowing $\geq$0.7~$R_\oplus$ planets to maintain their atmosphere. Initial carbon inventory emerges as the most influential parameter for atmospheric retention, though orders of magnitude difference to Earth values are required to make a significant difference to longevity of atmospheric retention. Planets with substantial initial carbon content, large amounts of heat producing elements, cool initial mantle temperatures and low core radius fractions show the best atmospheric retention capabilities. 
Our results indicate that atmospheric retention on small planets depends strongly on their formation conditions and early evolution, providing important constraints for future observations of rocky exoplanets and their potential habitability.

\end{abstract}

%%%%%%%%%%%%%%%%%%%%%%%%%%%%%%%%%%%%%%%%%%%%%%%%%%%%%%%%%%%%%%%%%%%%

\section{Introduction}
\label{intro}

The abundance of planet discoveries outside our Solar System enabled large statistical studies of exoplanets, providing insight into the formation of our system \citep{Howard2012,Dressing2013,Dressing2015,Kopparapu2013,bryson2021,kane2021b, Emsenhuber2021, Emsenhuber2025}. The plethora of exoplanets creates an interesting challenge in the search for potentially habitable planets: Of the many targets in the habitable zone (HZ) of their star, the region around a star where water can exist in a liquid state on the surface of a planet with sufficient atmospheric pressure \citep{kas93,Kopp13,Kopp14,Hill2023}, which are the best candidates for follow-up observations with the aim of detecting biosignatures? The occurrence rates of planets continue to increase towards smaller planet sizes \citep{Kane16}, and as the detection capabilities improve, the population of smaller than Earth planets will soon grow. Future missions such as PLATO \citep{rauer2014plato} have been designed to search for Earth-sized planets in the HZ of Sun-like stars. With limited observational resources, it is essential to refine the list of potentially habitable planets to those that are most likely to be habitable. While many studies have attempted to determine the size at which a rocky planet is more likely a sub-Neptune \citep{weiss2014,Rogers15,Wolfgang16,zeng2019,Luque2022}, the lower boundary of planet size for maintaining an atmosphere remains less explored. This paper fills this gap and will contribute to the strategic prioritization of targets in future exoplanet exploration missions.

As planet size reduces, its ability to maintain an atmosphere can be affected in many ways. Following the mass-radius relationship for terrestrial planets, as the size of the planet reduces so will the mass, depending on factors such as core-mass fraction \citep{Chen16,Unterborn2023,weiss2014,zeng2016a,otegi2020, Noack2020, unterborn2019}. A lower mass produces a lower surface gravity, allowing easier escape of atmospheric molecules due to a lower escape velocity, and thus atmospheric loss occurs faster \citep{Kislyakova2020, noack2017a}. Smaller planets also cool faster, shutting off volcanism and the associated degassing earlier \citep{foley2020}. Smaller planets will also have relatively cooler mantle temperatures over their lifetimes. 
Earth is currently the only planet known to have plate tectonics. All other rocky bodies in the solar system are currently thought to have a stagnant lid tectonic regime, though Venus' tectonic state remains debated and may represent something intermediate between stagnant lid and plate tectonics, and there is inconclusive evidence as to whether any solar system planets have had plate tectonics in the past \citep{Weller2019, Weller2023, Anderson1981}. For this reason this version of the model is based on a planet with a stagnant lid, where volatile cycling and mantle regassing is extremely limited. Stagnant lid planets maintain a largely immobile outer shell of lithosphere through which heat must be conducted from the interior. As planet size decreases, the reduced mass leads to lower internal pressures and generally cooler mantle temperatures, which reduce both the vigor of mantle convection and the planet's ability to maintain volcanic activity \citep{Valencia2006,FoleySmye18}. For smaller planets, the ratio of surface area to volume increases, resulting in more rapid cooling of the interior over time \citep{oosterloo2021}. This cooling rate directly impacts the longevity of volcanic outgassing by causing the lithosphere to thicken more rapidly relative to larger planets.
A thicker lithosphere inhibits heat transport from the interior and can effectively shut down surface volcanic activity, even when the mantle still contains significant volatile components \citep{noack2017a}. 
The initial composition and thermal state of these planets also play crucial roles in their evolutionary pathways. 
The abundance and distribution of radiogenic materials
provide a long-term heat source driving mantle convection and volcanic activity \citep{schubert2001}. In smaller planets, while the concentration of these materials may be similar to larger worlds, their total heat contribution is likely reduced due to the smaller planetary mass. This reduced heating, combined with more efficient cooling, can lead to early cessation of volcanic activity and limited replenishment of atmospheric gases \citep{FoleySmye18}. 
Atmosphere retention becomes increasingly challenging as planet size decreases due to lower surface gravity. 
Lower surface gravity in particular affects atmosphere loss through Jeans escape, a loss mechanism considered in this paper, which is an increasingly important method of loss for smaller planets \citep{Kislyakova2020}. 
In Jeans escape, high-energy molecules have sufficient kinetic energy to escape from the atmosphere into a nearly collision free exosphere \citep{Catling_Kasting_2017}. 
The Jeans escape parameter ($\lambda$) scales with the gravitational potential energy and inversely with atmospheric temperature at the exobase, the outermost layer of the planet's atmosphere. When 
$\lambda$ is large, particles are tightly bound and only the extreme high-velocity tail of the Maxwell–Boltzmann distribution can escape, giving very low escape rates. When $\lambda$ is small, thermal motions are equal or larger than the escape speed, leading to higher escape rates. For smaller planets, the reduced surface gravity leads to higher loss rates, as molecules need only reach relatively low velocities in order to escape.

\begin{figure*}
    \centering
    \includegraphics[width=1\linewidth]{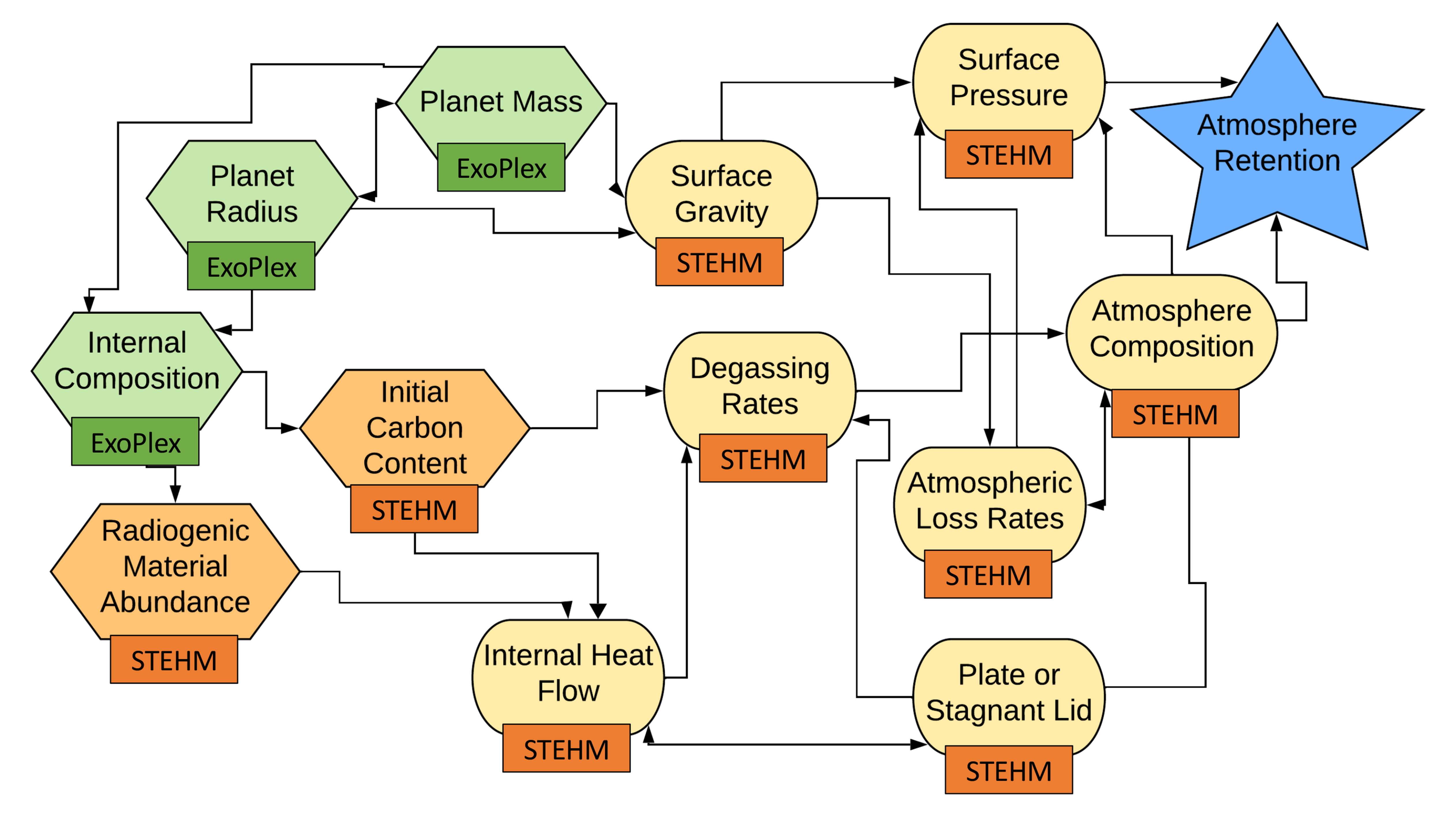}
    \caption{STEHM flow chart. Green hexagons are input parameters that are calculated by ExoPlex. Orange hexagons are input parameters set within STEHM. Yellow stadiums are components that are explored by STEHM. Arrows indicate how each section of the code interacts with the others. On the bottom right there is a yellow stadium indicating a choice of tectonic regime. This paper is based on a planet with stagnant lid tectonics, however future versions of the STEHM code will also include plate tectonics.  
}
    \label{fig:STEHM}
\end{figure*}

To investigate how the above processes effect the long term atmosphere retention of small planets we introduce the Smaller Than Earth Habitability Model (STEHM), a 1D model built on an interior thermal evolution and degassing model \citep{Foley2015,FoleyDris2016,FoleySmye18,Foley_2019}, with an atmosphere loss module based on the stellar flux estimates of a Sun-like star \citep{ribas2005} and corresponding CO$_2$ atmosphere loss rates \citep{tian2009a, tian2009b, Kite2020}. 
We use planet interior composition solver ExoPlex \citep{Unterborn2023} to determine the bulk mantle density,
mantle gravity and mass of the planets as we step down in size from 1.0~$R_\oplus$ to 0.5~$R_\oplus$. We model a pure CO$_2$ atmosphere, focusing on CO$_2$ as it is a heavy molecule, which combined with its radiative cooling effects, makes it a difficult molecule to lose \citep{tian2009a, tian2009b}. We focus on a CO$_2$ atmosphere to represent a best-case scenario for atmosphere retention.
A flow chart that shows how the modules of STEHM interact is shown in Figure \ref{fig:STEHM}. The green hexagons indicate input parameters that are calculated outside the code. Orange hexagons are input parameters set inside the code and yellow stadiums are components explored by the code. 
We systematically decrease the modeled planet size from 1~$R_\oplus$ to 0.5~$R_\oplus$ to identify the threshold at which the planet's atmosphere, a crucial criterion for habitability, ceases to exist. The results from STEHM will have implications for the selection and study of exoplanets in future space missions. 
In Section~\ref{Exoplex}, we describe ExoPlex and its use in determining some of the model inputs. Section~\ref{STEHM} presents the Smaller Than Earth Habitability Model. In Section~\ref{verify}, we confirm the validity of the model by testing its results against known Solar System stagnant lid planets. We explore a wide range of scenarios with STEHM and present the results in Section \ref{Results}. We discuss the result implications, caveats and proposed expansion opportunities in Section \ref{Disc}, and provide concluding remarks in Section~\ref{conclusions}.

%%%%%%%%%%%%%%%%%%%%%%%%%%%%%%%%%%%%%%%%%%%%%%%%%%%%%%%%%%%%%%%%%%%%

\section{Exoplex}
\label{Exoplex}

We use ExoPlex \citep{Unterborn2023} to determine the likely planet mass, bulk mantle density and mantle gravity of the planets as we step down in size from 1$~R_\oplus$ to 0.5$~R_\oplus$. The ExoPlex code iteratively solves for a planet’s density, pressure, gravity, and adiabatic temperature profiles that are consistent with the pressures derived from the mass within a sphere. Convergence is reached when the change of the difference in the density of all layers from the previous run to the
current is $<1\times10^{-6}$.
We set the default core radius fraction (CRF) to Earth's value (0.55) for each planet. Changes to this CRF are further explored in Section \ref{density}. 
We set the Exoplex depth slices for the core and mantle layers at 600 and 500, respectively. This sets the resolution of the mass-radius sampling, with too few layers resulting in insufficient resolution, while too many provide negligible improvement at a significant computational cost. Each layer is a radial shell with a small mass increment,
pressure, density, and gravity determined by the equation of state (EOS).
ExoPlex determines the gravity for each layer of the planet using Gauss’ Law of gravity, given the density and radius, then an average of the mantle and core layers determines their average gravity. The surface gravity of the planet is extracted from the outermost layer of the model. Mantle density is determined by stitching together the lower and upper mantle grids and interpolating within them to determine the
density of each layer in the planet, then taking an average of the layers.
The results from Exoplex are presented in Table~\ref{tab:ExoPlex}. These results were used as input parameters for STEHM.

\begin{deluxetable*}{lcccccc}
\tablecaption{\label{tab:ExoPlex} ExoPlex results for STEHM input parameters }
\tablehead{\colhead{Radius } & \colhead{Mass } &  
\colhead{Mantle Radius} & \colhead{Planet Density } & \colhead{Mantle Density } & \colhead{g$_{Mantle}$}  & \colhead{g$_{Surface}$} 
\\
\colhead{R$_\oplus$} & \colhead{M$_\oplus$} &  
\colhead{km} & \colhead{g~cm$^{-3}$} & \colhead{kg~m$^{-3}$} & \colhead{m~s$^{-2}$} & \colhead{m~s$^{-2}$}}
\startdata
1.0 & 1.00 &  
3198.54 & 5.49 & 4623 & 9.93 & 9.75 \\
0.9 & 0.69 &  
2854.96 & 5.21 & 4392 & 8.77 & 8.33 \\
0.8 & 0.46 &  
2521.52 & 4.95 & 4176 & 7.31 & 7.08 \\
0.7 & 0.29 &  
2195.08 & 4.69 & 3948 & 5.83 & 5.83 \\
0.6 & 0.17 &  
1872.40 & 4.40 & 3696 & 4.65 & 4.70 \\
0.5 & 0.09 &  
1552.26 & 4.16 & 3469 & 3.59 & 3.69 \\
\enddata
\end{deluxetable*}

%%%%%%%%%%%%%%%%%%%%%%%%%%%%%%%%%%%%%%%%%%%%%%%%%%%%%%%%%%%%%%%%%%%%

\section{The Smaller Than Earth Habitability Model}
\label{STEHM}

\subsection{Interior Thermal Evolution and Degassing Module}
\label{BradCode}

The thermal interior evolution and degassing module used in this paper is based on the stagnant lid planet model outlined by \citet{FoleySmye18, foley2019}. The heat loss from the mantle for the stagnant lid model is driven by conductive heat transfer through the lithosphere, which is governed by mantle convection and magmatism due to cooling of erupted magma to near-surface temperatures, and the latent heat released during solidification. The rate of heat conduction is controlled by the thickness and thermal properties of the lithosphere. The model calculates how the thickness of this lithosphere evolves over time based on the temperature difference between the mantle and surface, taking into account both conductive heat transfer and internal heat generation from heat producing elements (HPE). Rather than modeling core cooling, the model assumes pure internal heating, which provides a conservative estimate for how quickly a planet cools and volcanic activity ceases. This provides an estimate of the shortest degassing timeline as including core cooling would prolong the time that volcanism would occur, and would add more poorly constrained parameters to the model \citep{FoleySmye18}. 
The model handles mantle melting and crustal production by considering how mantle material melts as it upwells and decompresses. When upwelling mantle crosses its melting temperature, partial melting occurs, with the amount of melt depending on how far above the melting temperature the mantle reaches. This melt then contributes to crustal formation at the surface \citep{FoleySmye18}.  
The model employs a simplified carbon cycle where CO$_2$ moves between two reservoirs: the mantle and the atmosphere. The model releases CO$_2$ through volcanic outgassing from mantle-derived melts  
and then accumulates in the atmosphere. The model assumes all melt produced travels to the surface or near surface, and that all melt contributes to degassing of CO$_2$ to the atmosphere.
The time at which CO$_2$ outgassing stops is primarily driven by the cooling of the planet's mantle. 
A primary source of internal heat comes from the decay of radioactive isotopes, particularly potassium-40 (K), thorium-232 (Th), uranium-235 (U$_{235}$) and uranium-238 (U$_{238}$). As these elements decay over billions of years, their heat production gradually diminishes. The model accounts for the changing contributions of each isotope based on their different decay rates and initial abundances. 
As HPE are incompatible elements they are preferentially lost from the solid to the melt, causing the crust to become enriched in HPE over time. All crust buried to depths below
the lithospheric thickness founders into the mantle, recycling HPE back into the mantle.

\citet{FoleySmye18,foley2019} demonstrated that the main contributors to significant change in the model results were the initial size of the carbon mantle reservoir, initial volume of HPE, and the initial mantle potential temperature. For each of these parameters we have default values outlined below and in Table~\ref{tab:Variable}. To thoroughly examine the parameter space, we run the model using the minimum and maximum values within specified ranges taken from the literature for these parameters, also outlined in Table \ref{tab:Variable}.

\begin{deluxetable*}{lcccc}
\tablecaption{\label{tab:Variable} STEHM Variable Parameters. }
\tablehead{\colhead{Parameter } & \colhead{Units } &  
\colhead{Min} & \colhead{Max } & \colhead{Default } }
\startdata
Initial carbon budget & mols & 7$\times10^{21}$ & 2$\times10^{22}$ & 1.6$\times10^{22}$ \\
 & mols & 1$\times10^{21}$ & 1$\times10^{23}$ &  \\
Mantle initial temp & K & 1500 & 2200 & 1900 \\
U238 & ppm & 0.45$\times$Default & 1.92$\times$Default & 0.022$\times$0.9927 \\
U235 & ppm & 0.45$\times$Default & 1.92$\times$Default & 0.022$\times$0.0072 \\
Th & ppm & 0.77$\times$Default & 1.88$\times$Default & 0.083 \\
K & ppm & 0.35$\times$Default & 3.63$\times$Default & 261(0.0117/100) \\
Core Radius Fraction &  & 0 & 0.7 & 0.55 \\
Distance from Star & AU & 0.75 & 1.765 & 1 \\
Exobase Temperature & K & 800 & 2500 & 400
\enddata
\end{deluxetable*}

The initial mantle carbon budget default value is $1.6\times10^{22}$ moles of carbon. This value was determined through our calibration outlined in Section \ref{verify} and is within the \citet{sleep2001b} estimate of 7$\times10^{21}$ to 2$\times10^{22}$ mole of CO$_2$ in Earth's mantle. When testing the effect that the variation of carbon budget has on model results, we use the upper and lower limits of the \citet{sleep2001b} estimates. We also test a wider range of 1$\times10^{21}$ to 1$\times10^{23}$ mole of CO$_2$ to fully explore how initial carbon budget affects the long term ability of smaller than Earth sized planets to maintain an atmosphere.

The default values for initial HPE budget (K, Th, U$_{235}$, U$_{238}$) are from \citet{Palme2003}. We refer to \citet{Unterborn2023} Appendix A4 for the minimum and maximum range of HPE concentrations.
They calculated a current-day ($t = 12.5$~Gyr after the birth of the Milky Way) range of HPE abundance with a 95\% confidence interval of (Th=Mg)$_{star}$ to be between 0.77--1.88 $\times$ Solar, Europium (as a proxy for Uranium) (Eu/Mg)$_{star}$ = (U/Mg)$_{star}$ ranges between 0.45--1.92 $\times$ Solar, and the Potassium (K/Mg)$_{star}$ range falling between 0.35--3.63 $\times$ Solar.

The default initial mantle temperature is 1900~K. When testing the effect that variation of initial mantle temperature we use the range of T$_i$=1500--2200~K, set out in \citet{foley2019}.

We use the ExoPlex values from Table~\ref{tab:ExoPlex} to set the mantle density and gravity. 
The surface temperature is fixed at 288~K, as it was shown in \citet{foley2019} that surface temperature changes of the order of 100~K do not significantly impact the results of the model. The variable parameters of the model are set out in Table~\ref{tab:Variable}, and the remaining default model values for all model input parameters are taken from \citet{foley2019}, Table 1. A visual of how our model components interact is shown in Figure \ref{fig:STEHM}.

%%%%%%%%%%%%%%%%%%%%%%%%%%%%%%%%%%%%%%%%%%%%%%%%%%%%%%%%%%%%%%%%%%%%

\subsection{Atmosphere Escape Module}
\label{Atmesc}

\begin{figure}
\begin{center}
\includegraphics[width=0.5\textwidth]{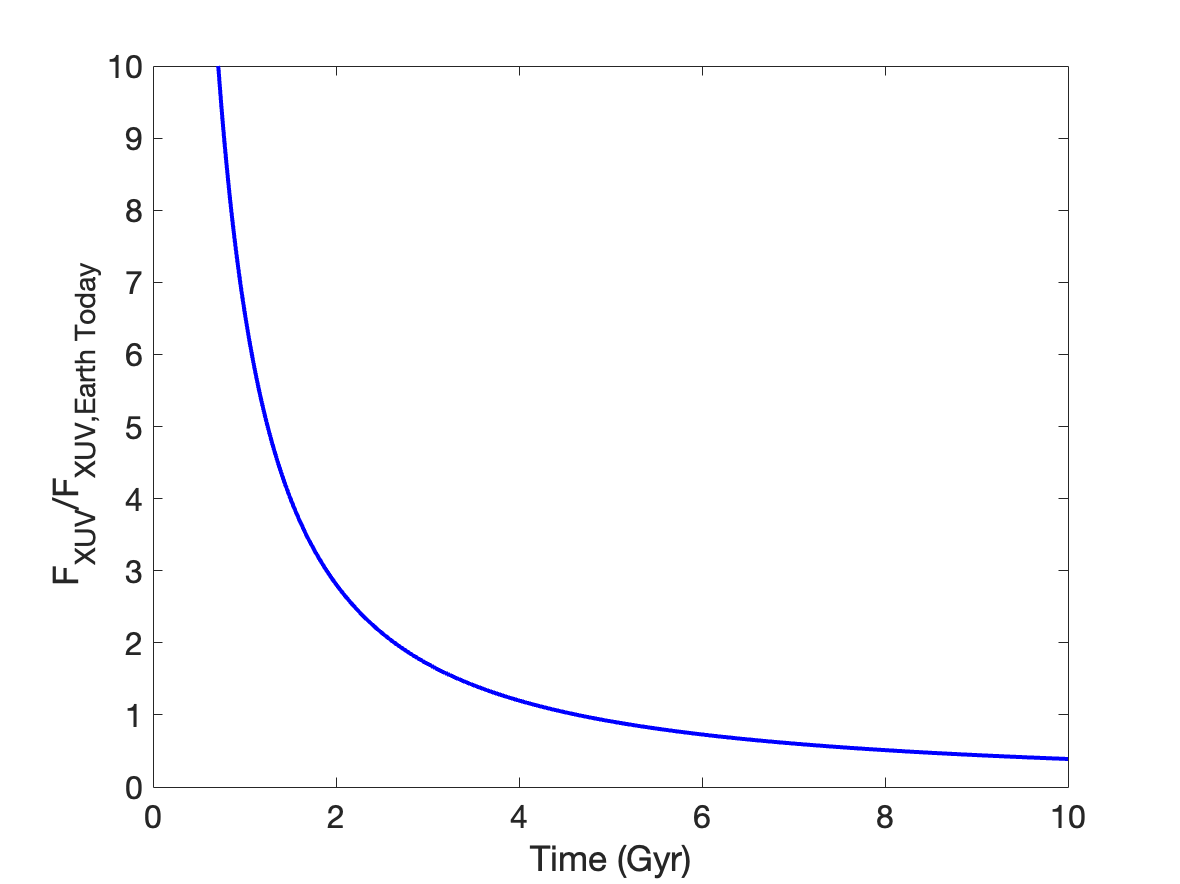}
  \end{center}
\caption{Flux rates of a Sun-like star from \citet{ribas2005}, normalized so that the flux at the present solar age matches the modern solar XUV value. We adopt a conservative maximum stellar XUV flux of 10 times the modern solar value. 
\label{fig:Ribas}}
\end{figure}

The atmosphere loss module is based on the stellar flux estimates of a Sun-like star \citep{ribas2005} (Figure \ref{fig:Ribas}). 
We apply a uniform normalization factor so that the flux at 4.6 Gyr agrees with the present-day solar XUV flux. The unscaled relation over-predicts the modern solar XUV flux by a factor of 4.5, and so we reduce the flux at all ages by this same factor. This correction preserves the original evolution with time of the \citet{ribas2005} relation while recalibrating it to the present-day Sun. We also adopt a conservative maximum stellar XUV flux of 10x current XUV flux levels for the first 700~Myrs, after which the normalized \citet{ribas2005} power law takes effect. While \citet{ribas2005} estimated the young Sun emitted XUV radiation at levels up to $\simeq$100x the present-day value during its early evolution, subsequent studies have shown that this may be an over estimation, with lower fluxes (10–50x) more plausible for solar analogs \citep{Johnstone2015,Tu2015}. Early XUV flux rates are dependent on the rotational history of the star, with slow rotators exhibiting lower XUV flux levels. Models of stellar spin evolution indicate that the Sun’s current rotation is most consistent with slow to moderate rotational evolutionary tracks, suggesting that the early Sun may have experienced a lower XUV flux history \citep{Tu2015, gallet2015}.

\begin{figure}
\begin{center}
\includegraphics[width=1\columnwidth]{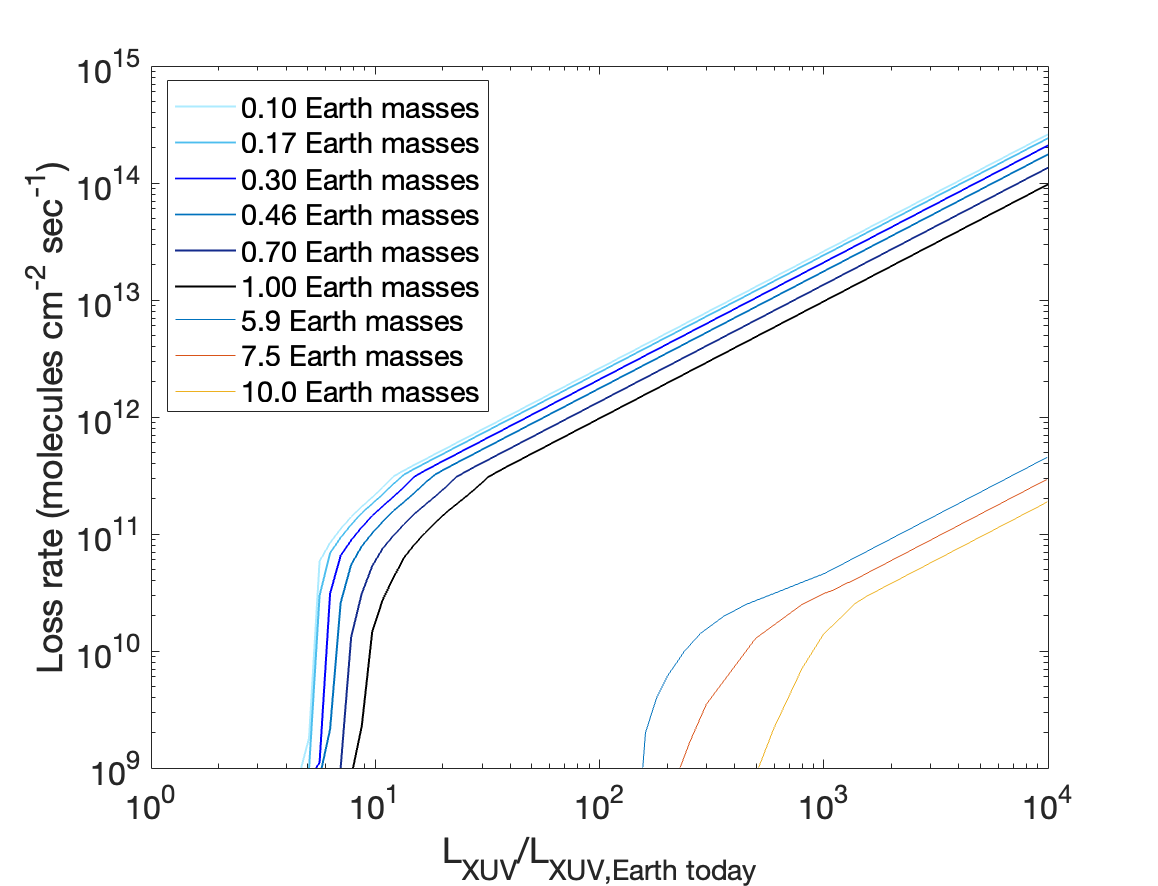}%
\end{center}
\caption{Atmosphere loss rates for the planets tested in this model using the masses calculated by ExoPlex (Table \ref{tab:ExoPlex}). Loss rate in the model drops to zero once rate reduces to $10^9~cm^{-2}s^{-1}$. Loss rates are determined from those provided in \citet{Kite2020} Figure S4, consistent with the adopted figure template. The 5.9, 7.5 and 10.0 Earth mass loss rates from \citet{Kite2020} are shown for comparison. The 0.10 Earth-mass case from \citet{Kite2020} coincides with our model results.   
\label{fig:atm_loss}}
\end{figure}

Atmospheric loss of CO$_2$ is based on the 1D hydrodynamic planetary thermosphere-ionosphere model from \citet{tian2009a,tian2009b} in which an electron transport-energy deposition model is coupled to the thermosphere-ionosphere model. It was used to model the Martian thermosphere-ionosphere to study CO$_2$ escape from early Mars \citep{tian2009a} as well as super-Earth planets in the HZ of M-dwarf stars \citep{tian2009b}. The Mars escape rates were scaled by \citet{Kite2020} to Earth's orbital distance of 1~AU (F$_{XUV}~\propto~1/S^2$), and we adopt these values (Figure \ref{fig:atm_loss}). 

The \citet{tian2009a} model dynamically determined the exobase altitude and adjusted its upper boundary accordingly. At this upper limit, it employed the thermal atmosphere loss mechanism of Jeans escape to calculate effusion velocities for different atmospheric species when XUV flux rates were between 5-9.99$\times$ present day solar XUV rates. The temperature at the exobase causes air particles to have an approximately Maxwellian velocity distribution. Jeans escape occurs when the particle velocity at the upper end of the distribution exceeds the escape velocity of the planet
[\citep{jeans1925}. Provided the particle is in the exosphere, or outer most layer of the atmosphere, where there are little to no other particles to collide with, the particle will escape to space.  
The \citet{tian2009a} model included dissociation of CO$_2$ under high XUV levels, and the escaping material was atomic C and O. 
At XUV flux rates $\geq$10$\times$ present day solar XUV rates the atmosphere loss entered the hydrodynamic loss regime. While in this regime a planet would suffer great atmosphere loss with the bulk motion velocity of the atmosphere increasing to escape velocity and mass outflow facilitating the loss of heavier molecules CO and CO$_2$ through atmospheric drag.
The model also included radiative cooling. In CO$_2$ dominated atmospheres, radiative cooling channels include the CO$_2$ 15-$\mu$m band emission, O emission at 63-$\mu$m, and the CO 4.7-$\mu$m band. The combined effect of these cooling mechanisms can significantly reduce atmospheric escape rates compared to atmospheres lacking efficient coolants, though under high XUV flux the molecular species are dissociated, reducing cooling efficiency.

We focus on a pure CO$_2$ atmosphere as the best-case scenario for atmosphere retention as the heavy molecule is more difficult to lose through Jeans escape mechanisms \citep{dong2020}, which combined with the radiative cooling effects of CO$_2$ can significantly reduce atmospheric escape rates. 
We use a default value of $T = 400$~K for the atmosphere temperature at the exobase, and test exobase temperatures of 800~K and 2500~K, as used by \citet{tian2009a}. We validated the atmosphere escape model by recreating the results from \citet{Kite2020} fig S4, and show the atmosphere loss rates for the planets tested in Figure~\ref{fig:atm_loss}. The atmosphere loss rates are set to drop to zero once the rate reduces to $10^9$~cm$^{-2}$s$^{-1}$.

\begin{figure}
\begin{center}
\includegraphics[width=0.5\textwidth]{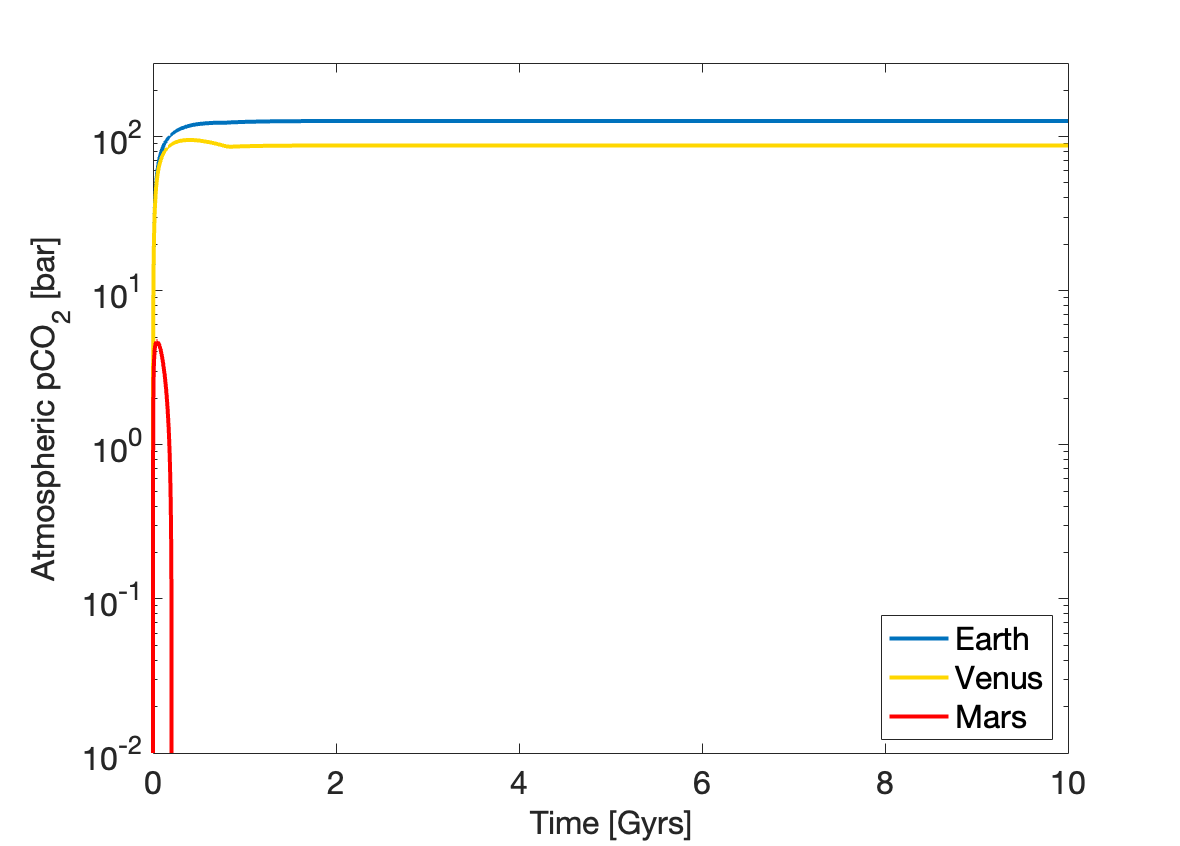}
  \end{center}
\caption{STEHM results for stagnant lid Earth (Blue), Venus (Yellow) and Mars (Red) analogs. The model predicts Earth as a stagnant lid planet could have gained a 126~bar CO$_2$ atmosphere,  
and Venus an 88 bar CO$_2$ atmosphere. After a early degassing of over 4~bars of CO$_2$, Mars' atmosphere is depleted by $\sim200$~Myrs.
\label{fig:EarthVenusMars}}
\end{figure}

%%%%%%%%%%%%%%%%%%%%%%%%%%%%%%%%%%%%%%%%%%%%%%%%%%%%%%%%%%%%%%%%%%%%

\section{Verification of Model Against Solar System Planets}
\label{verify}

Venus and Mars are used as observational benchmarks to constrain the model parameter space. We require that the model reproduce their first-order atmospheric evolutions of a dense ($\sim90$ bar), long-lived CO$_2$ atmosphere for Venus and a transient, multi-bar CO$_2$ atmosphere for early Mars that is subsequently lost \citep{Kite2014,edwards2015,jakosky2018,Thomas2023,hamano2013,Turbet2021}. These cases provide end-member examples of stagnant-lid planetary evolution within the Solar System.

The initial mantle carbon inventory is not well constrained and is typically estimated to lie within
$\sim 10^{22} - 10^{23}$ mol for terrestrial planets \citep{sleep2001b,DASGUPTA20101,marty2012}. We therefore treat it as a free parameter within this physically plausible range and calibrate it using the Venus and Mars benchmarks set out earlier. We find that an initial mantle inventory of $1.6\times10^{22}$ mol C, scaled to the mantle mass of the planet compared to Earth's mantle mass, is the minimum value that satisfies both constraints. Lower values fail to generate sufficient atmospheric CO$_2$ for the Venus case, while higher values are not required. We therefore adopt $1.6\times10^{22}$ mol as an observationally calibrated estimate of the degassable carbon inventory for the default version of the model.

We calculate the flux received on Venus and Mars at their respective distances from the sun so that their atmosphere loss rates can be adjusted in the model accordingly. We find with an insolation scaling of $1/S^2$ that at 0.72~AU Venus receives 1.93~$F_\oplus$ and 
at 1.5~AU Mars receives or 0.45$F_\oplus$. 

We also include a stagnant-lid-Earth scenario to test how such a planet would evolve and found it would have gained a 126~bar CO$_2$ atmosphere. Note that while this seems to imply that such a planet at Earth’s current HZ position would be dominated by a dense CO$_2$ envelope and potentially uninhabitable, our model does not include CO$_2$ draw down through weathering as our focus is on maximizing atmosphere retention. Provided water was available, \citet{FoleySmye18} found that 0.01--1$\times$Earth's carbon inventory could be maintained at habitable conditions through weathering of CO$_2$ with fresh basalt produced via volcanism, and so these stagnant lid planets may not necessarily have thick CO$_2$ atmospheres.

%%%%%%%%%%%%%%%%%%%%%%%%%%%%%%%%%%%%%%%%%%%%%%%%%%%%%%%%%%%%%%%%%%%%

\subsection{Mars}
\label{Mars}

We set the model parameters using Mars' estimated initial carbon and HPE budgets to test the model results would align with Mars' expected history.  
We refer to \citet{YOSHIZAKI_2020} for the ranges of Mars' HPE concentrations and carbon content. Tables 3 and 5 list the bulk silicate Mars (BSM) values as K = 360 ppm, Th = 68 ppb, U = 18 ppb, and C =$\sim32$~ppm. \citet{YOSHIZAKI_2020} note that these estimates have large uncertainties, given the degree of mantle degassing is essentially unconstrained. We use the values for Mars mass from Table~7 of \citet{YOSHIZAKI_2020} and, assuming a uniform distribution of carbon in bulk silicate Mars, determine the mantle contains approximately $1.33~\times~10^{21}$~moles of Carbon. We use a mantle density of $\rho$=3640~kg~m$^{-3}$.

As described in Section~\ref{verify} we determine the flux received by Mars and calculated the corresponding atmosphere loss. We find that the Mars analog planet degases over 4 bars of CO$_2$ into the atmosphere before it is quickly lost again at $\sim200$~Myr.  
Our results agree with other studies that predict that young Mars likely had a thick early CO$_2$ rich atmosphere but would not have been able to maintain the atmosphere due to the high XUV from the early sun 
\citep{tian2009a,Scherf2020,Kite2014,jakosky2018,jakosky2023}.

%%%%%%%%%%%%%%%%%%%%%%%%%%%%%%%%%%%%%%%%%%%%%%%%%%%%%%%%%%%%%%%%%%%%

\subsection{Venus}
\label{Venus}

We adopted input values representative of Venus to verify that the model results would align with Venus' expected history. We use the mantle mass and density estimates provided by \citet{dumoulin2017}, who found an upper mantle density range of 3300--3400 kg m$^{-3}$, and lower mantle density range of 4800--5200 kg m$^{-3}$. Based on Earth's structure we assume the upper mantle comprises about 30\% of the total mantle volume, and the lower mantle 70\%. 
We determine the weighted average mantle bulk density = 4505 kg m$^{-3}$.
For initial carbon content we use the default value of $1.6\times10^{22}$ mol, scaled to the size of the Venusian Mantle. 
It was discussed in \citet{Nimmo2002} that Venus' heat production rate is likely similar to Earth's, implying similar HPE concentrations. Thus for initial HPE content we use Earth values scaled to the size of the Venusian Mantle compared to Earth's. The model predicts Venus quickly gains 93 bar atmosphere, then as degassing rates reduce it loses some atmospheric pressure until it maintains a steady atmospheric pressure of $\sim$88~bar for the length of the model run (Figure~\ref{fig:EarthVenusMars}). 

We also tested Venus and Mars analogs at 1~AU. The Venus planet was able to maintain a 104 bar atmosphere due to the lower flux received, the Mars planet initially degassed $\sim$3~bar of CO$
_2$ but lost its atmosphere completely by 80~Myrs.

%%%%%%%%%%%%%%%%%%%%%%%%%%%%%%%%%%%%%%%%%%%%%%%%%%%%%%%%%%%%%%%%%%%%

\section{STEHM Results}
\label{Results}

For all model runs we adopt the default values outlined in Section~\ref{BradCode} unless otherwise specified. We start at 1~R$_\oplus$ and step down to 0.5~R$_\oplus$  in increments of 0.1~R$_\oplus$. To thoroughly examine the parameter space, we also run the model using the minimum and maximum values within the specified ranges extracted from the literature for the initial size of the carbon mantle reservoir, initial volume of HPE, initial mantle potential temperature, planet density, and position of planet in the HZ.

\setlength{\belowcaptionskip}{2pt}
\vspace{0.00mm}
\begin{figure*}
\centering
\subfloat{%
  \includegraphics[width=1\columnwidth]{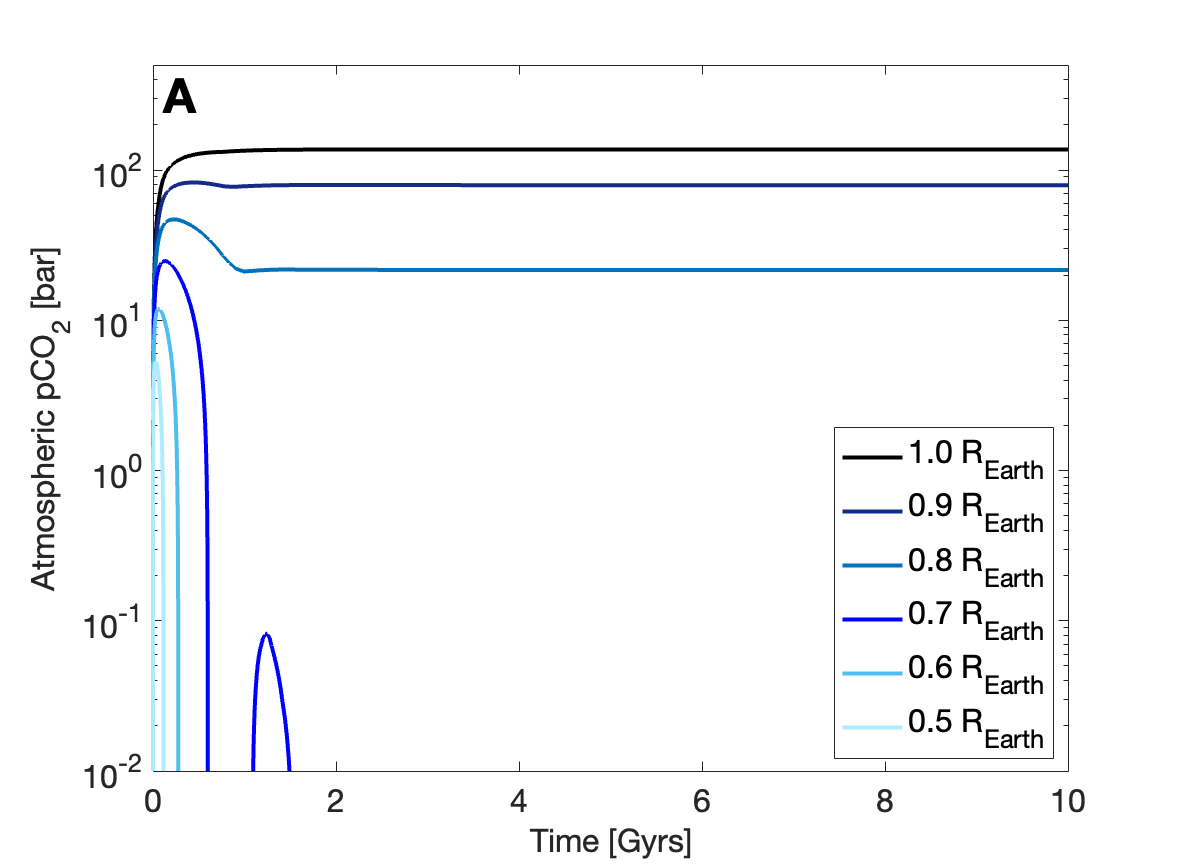}%
}
\vspace*{-5mm}
\subfloat{%
  \includegraphics[width=1\columnwidth]{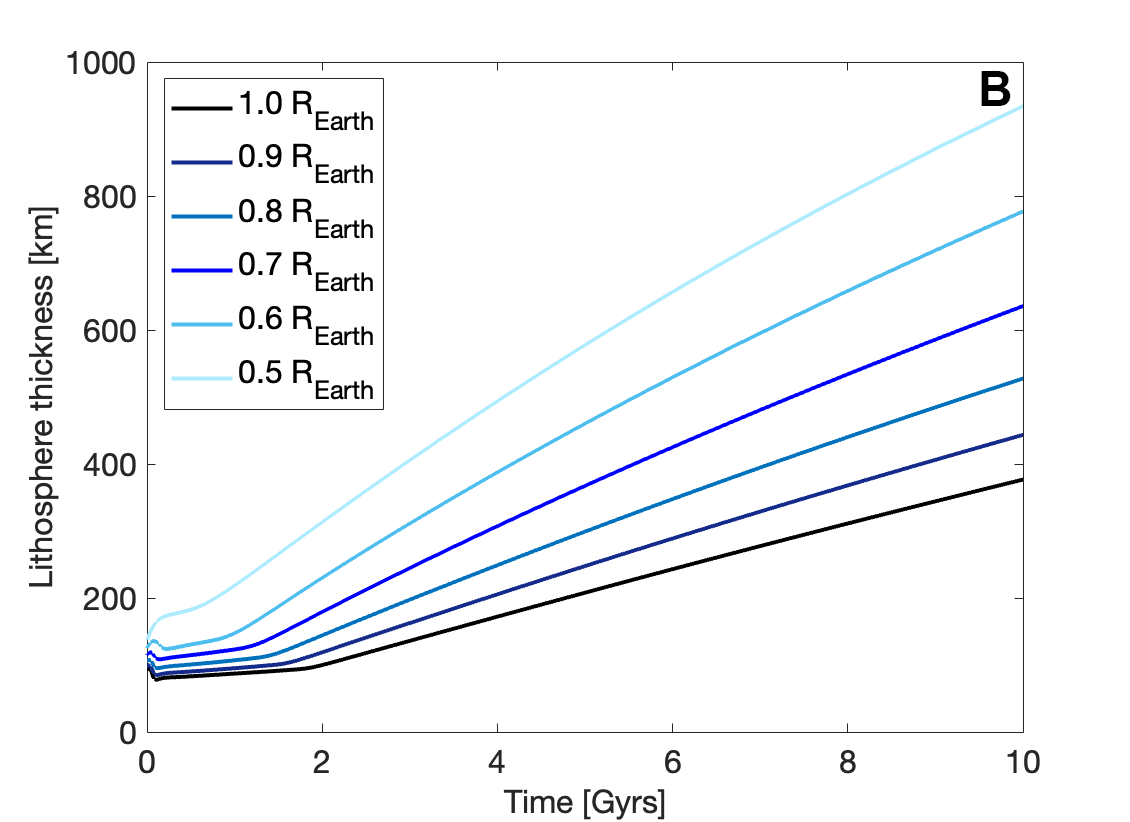}%
}\qquad
\vspace*{-5mm}
\subfloat{%
  \includegraphics[width=1\columnwidth]{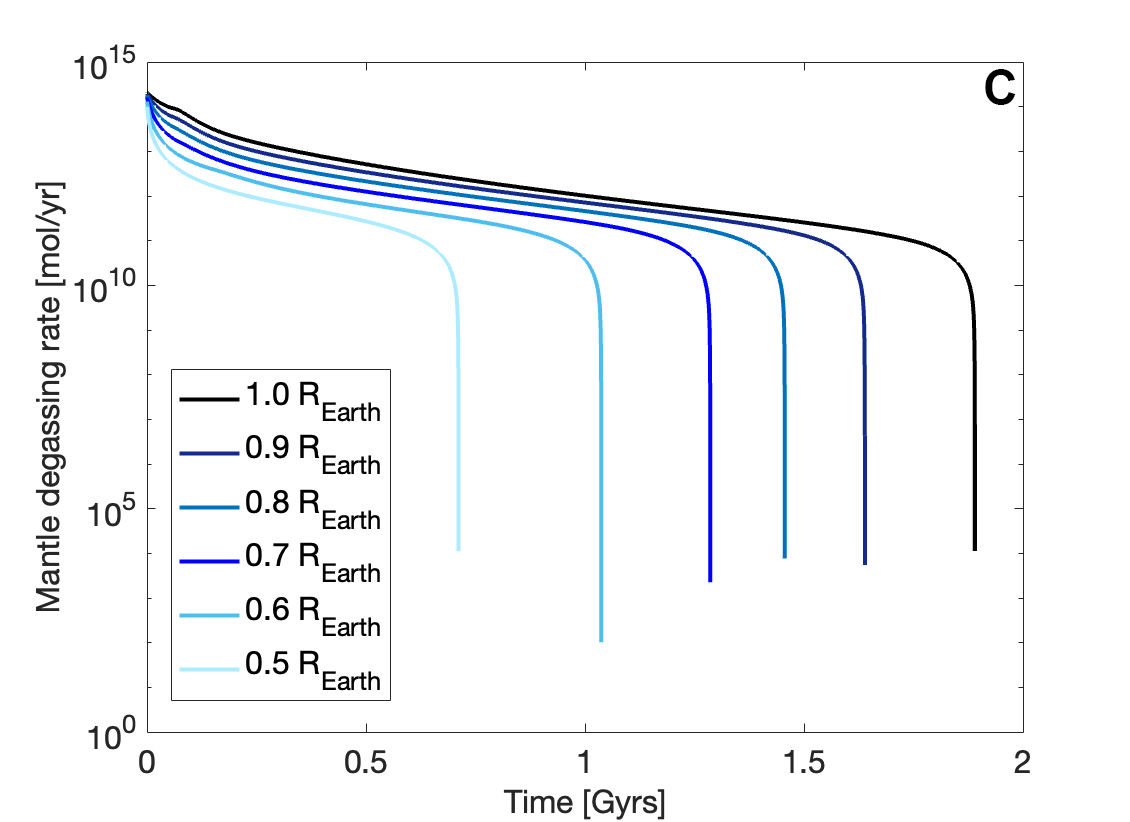}%
}
\vspace*{-5mm}
\subfloat{%
  \includegraphics[width=1\columnwidth]{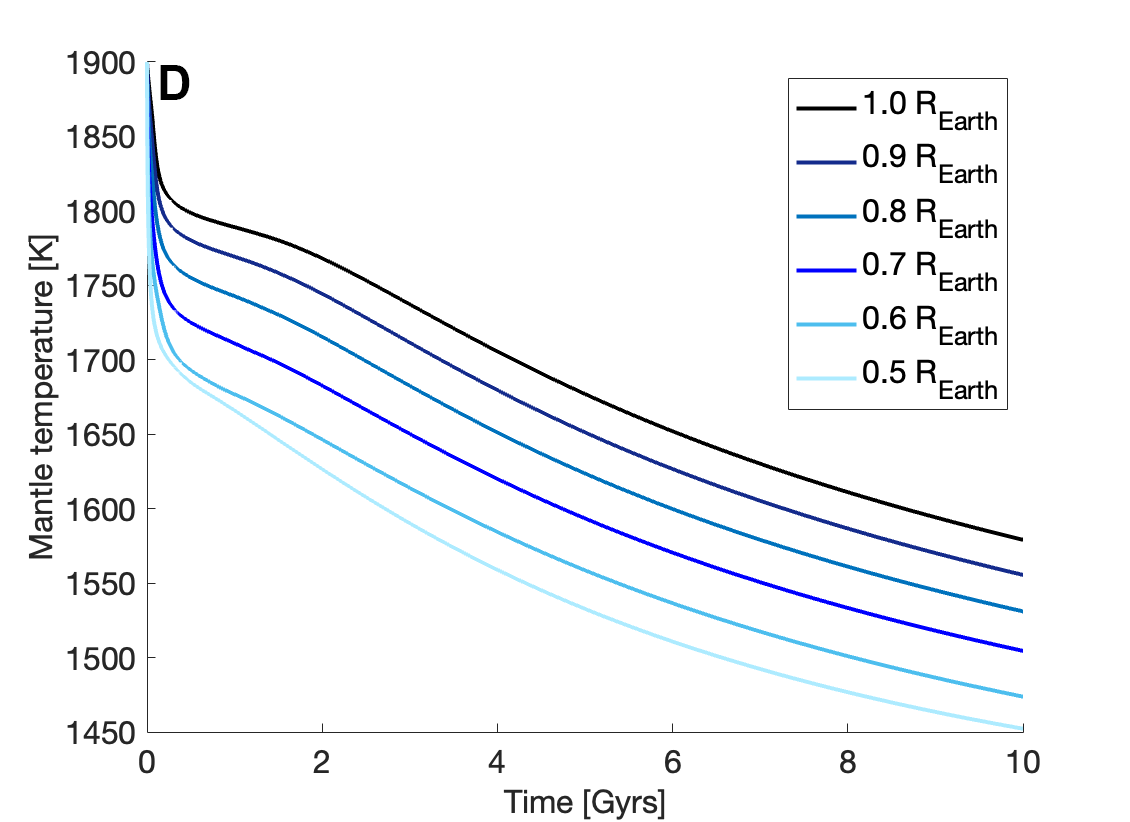}%
}
\vspace*{10mm}
\caption{Default Values. Planet size ranges from 1.0 - 0.5~R$_\oplus$. Panel A shows STEHM results for amount of atmospheric CO$_2$ found on a Earth sized and below planet. The model predicts that a stagnant lid planet would need to be $\geq$ 0.8~R$_\oplus$ to continuously maintain an atmosphere. Each of the 0.7, 0.6 and 0.5 R$_\oplus$ planets lose their atmospheres, though the 0.7~R$_\oplus$ planet has a short lived recovery of 0.08~bar. Panel B shows the lithosphere growth over time, with growth of the larger 1.0~R$_\oplus$ planet occurring much more gradually than the 0.5~R$_\oplus$ planet. This relatively fast growth contributes to the early cessation of the planet degassing for the smaller planets. Panel C shows the degassing rate for each planet size. The 1.0~R$_\oplus$ mass planet degases for the longest length of time, as well as degases more at each point in time. In Panel D we see the change in mantle temperature over time, with the larger planets maintaining a higher mantle temperature over their lifetimes. 
\label{fig:default_everything}}
\end{figure*}

The results of the default model run indicate a stagnant lid planet around a Sun-like star at Earth distance would need to be 
$\geq$~0.8~R$_\oplus$ to maintain an atmosphere past 1 billion years (Figure \ref{fig:default_everything}). Panel A of Figure \ref{fig:default_everything} shows STEHM results for the amount
of atmospheric pressure. The 1.0, 0.9 and 0.8~R$_\oplus$ planets maintain atmospheres of 137, 79, and 22~bar respectively. 
The 0.7~R$_\oplus$ planet loses its atmosphere at $\sim 0.6$~Gyrs, and then briefly regains a tenuous 0.08~bar atmosphere once the solar flux reduces to a point where CO$_2$ can rebuild in the atmosphere. The atmosphere is then lost again once degassing of the planet reduces and eventually shuts off (Figure \ref{fig:default_everything}, Panel C). Subsequently, smaller planets have their atmospheres depleted even earlier, with the 0.6~R$_\oplus$ planet losing its atmosphere at $\sim 0.4$~Gyrs, and the 0.5~R$_\oplus$ planet losing its atmosphere at $\sim 0.03$~Gyrs.  
Note the variation in results for the 0.5~R$_\oplus$ planet compared to the Mars analog is due to the change in distance of the planet from the star, along with changes in using Mars predicted values for mantle density, carbon content, and HPE content to using the ExoPlex values for the parameters outlined in Table~\ref{tab:ExoPlex}, and default Earth values scaled down to the size of the planet. 

Panel B of Figure~\ref{fig:default_everything} shows the  
growth of the lithosphere thickness.
The degassing slow down and shutoff is primarily driven by the cooling of the mantle and growth of the lithosphere thickness. Once the melt production stops, degassing ceases. It is expected that long-term planetary degassing will still occur through hot spots, but this type of degassing is not included in the model. Figure~\ref{fig:default_everything}, Panel C shows the degassing rate for each planet size. The 1.0~R$_\oplus$ mass planet degasses for the longest length of time, as well as degasses more over all at each point in time. This is driven partly by the ability of the larger planets to maintain a higher mantle temperature over their lifetime as seen in Figure~\ref{fig:default_everything}, Panel D.

\subsection{Variations in Initial Carbon Abundance}
\label{carbon}

We investigated the model's sensitivity to variations in initial planetary carbon content while maintaining all other parameters at their default values. The results reveal a strong dependence of atmospheric retention on both planetary size and initial carbon inventory. 
We run the model with the minimum value for initial carbon mantle budget on Earth as estimated by \citet{sleep2001b} of $7\times10^{21}$ and find significantly lower CO$_2$ in the atmosphere of each planet size. Under these conditions, only planets $\geq0.9~R_\oplus$ maintain their atmospheres, with the 1.0 and 0.9~$R_\oplus$ planets holding onto a 55 and 18~bar atmospheres respectively. The planets with radii $\leq0.8~R_\oplus$ all lose their atmospheres, though the 0.8~$R_\oplus$ planet recovers a 0.15~bar atmosphere that slowly diminishes over 3~Gyrs and then remains steady at 0.011~bar. Using the maximum value for carbon of $2\times10^{22}$ from \citet{sleep2001b} we find that the $\geq0.8~R_\oplus$ planets are able to maintain an atmosphere, with the 1.0, 0.9 and 0.8~$R_\oplus$ planets holding onto a 173, 106 and 40~bar atmosphere respectively. The planets with $\leq0.7~R_\oplus$ lose their atmosphere, though the $0.7~R_\oplus$ planet again has a brief recovery of 0.12~bar.

To explore a wider range of theoretical exoplanet scenarios, we also modeled cases with an initial carbon content an order of magnitude higher and lower than Earth's estimated value. With a carbon inventory of $1\times10^{23}$ moles, nearly all modeled planets retained their atmospheres throughout the simulation period, with only the $0.5~R_\oplus$ planet unable to hold its atmosphere past 1~Gyr. In contrast, when the initial carbon content is reduced to $1\times10^{21}$ moles, only the $1.0~R_\oplus$ planet is able to maintain an atmosphere of 0.8~bar. None of the planets $\leq0.9~R_\oplus$ maintain their atmospheres. These results underscore the critical role of initial carbon inventory in determining a planet's long-term atmospheric stability and highlight the importance of considering volatile inventories when assessing potential planetary habitability.

\begin{figure*}
\centering
\subfloat{%
  \includegraphics[width=1\columnwidth, height=0.8\columnwidth]{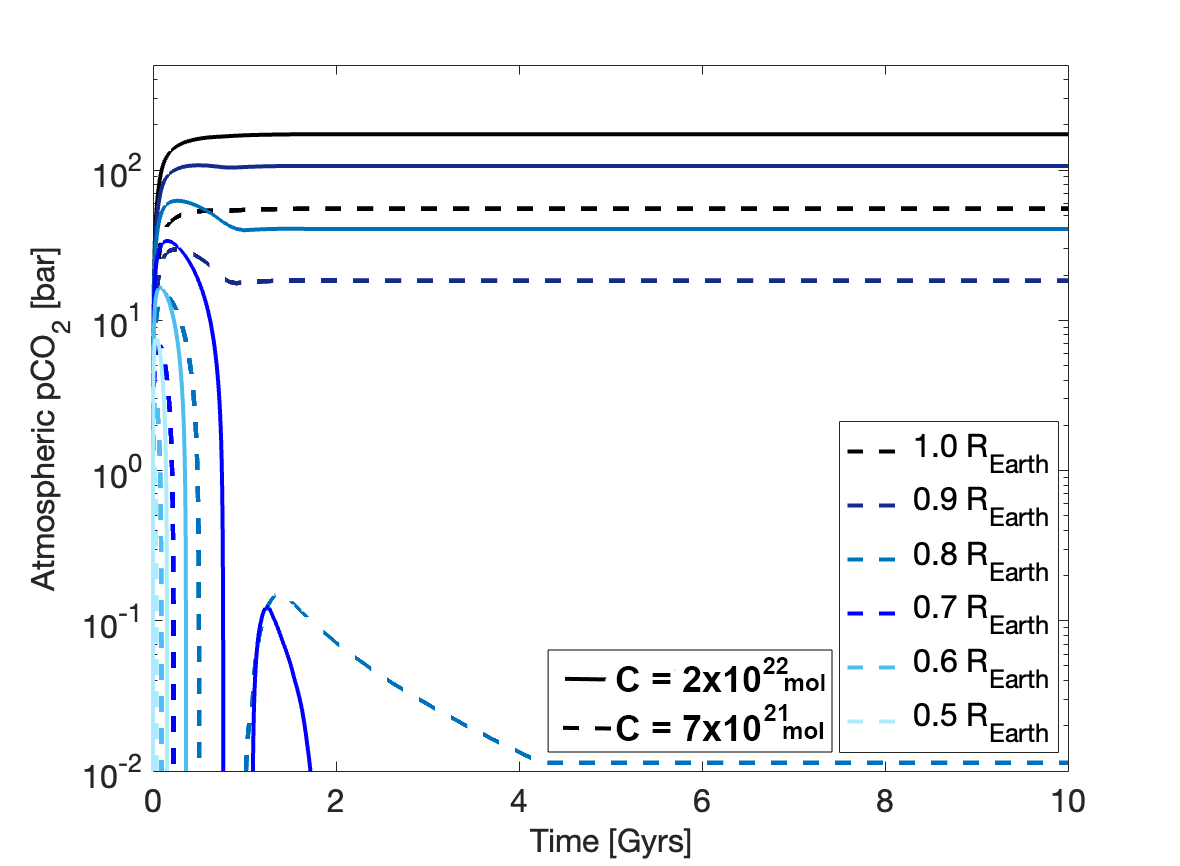}%
}
\subfloat{%
  \includegraphics[width=1\columnwidth, height=0.8\columnwidth]{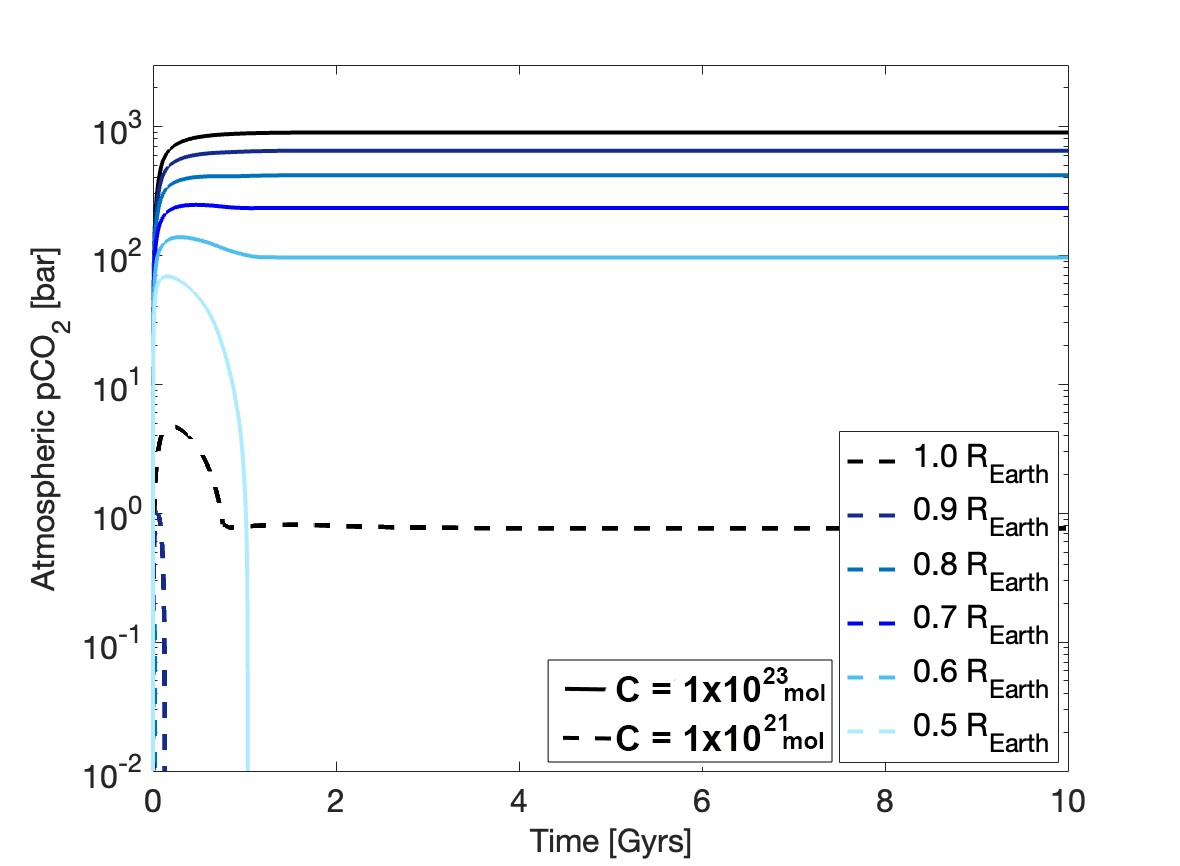}%
}
\caption{Variation in Initial Carbon Inventory. Other than the initial carbon content, the model uses default values set out in Section \ref{BradCode} for the remaining variables. Left: Model results using a maximum initial carbon content of $2\times10^{22}$ moles is represented by the solid lines and the minimum carbon content value of $7\times10^{21}$ moles is shown by the dashed lines. The legend shows the color representative of each planet radius. For the maximum initial carbon model runs, all planet $\geq0.8~R_\oplus$ maintain their atmosphere. With the minimum amount of initial carbon content, only the planets $\geq0.9~R_\oplus$ maintained their atmosphere. Right: Model results using a maximum initial carbon content of $1\times10^{23}$ moles is represented by the solid lines and the minimum carbon content value of $1\times10^{21}$ moles is shown by the dashed lines. Planets $\geq0.6~R_\oplus$ with high initial carbon content all maintain their atmospheres for the length of the model, while only the $1.0~R_\oplus$ planet with low initial carbon content maintained theirs.
\label{fig:carbon_vary}}
\end{figure*}

\subsection{Variations in HPE Abundance}
\label{hpe}

\begin{figure*}
\centering
\subfloat{%
  \includegraphics[width=1\columnwidth, height=0.8\columnwidth]{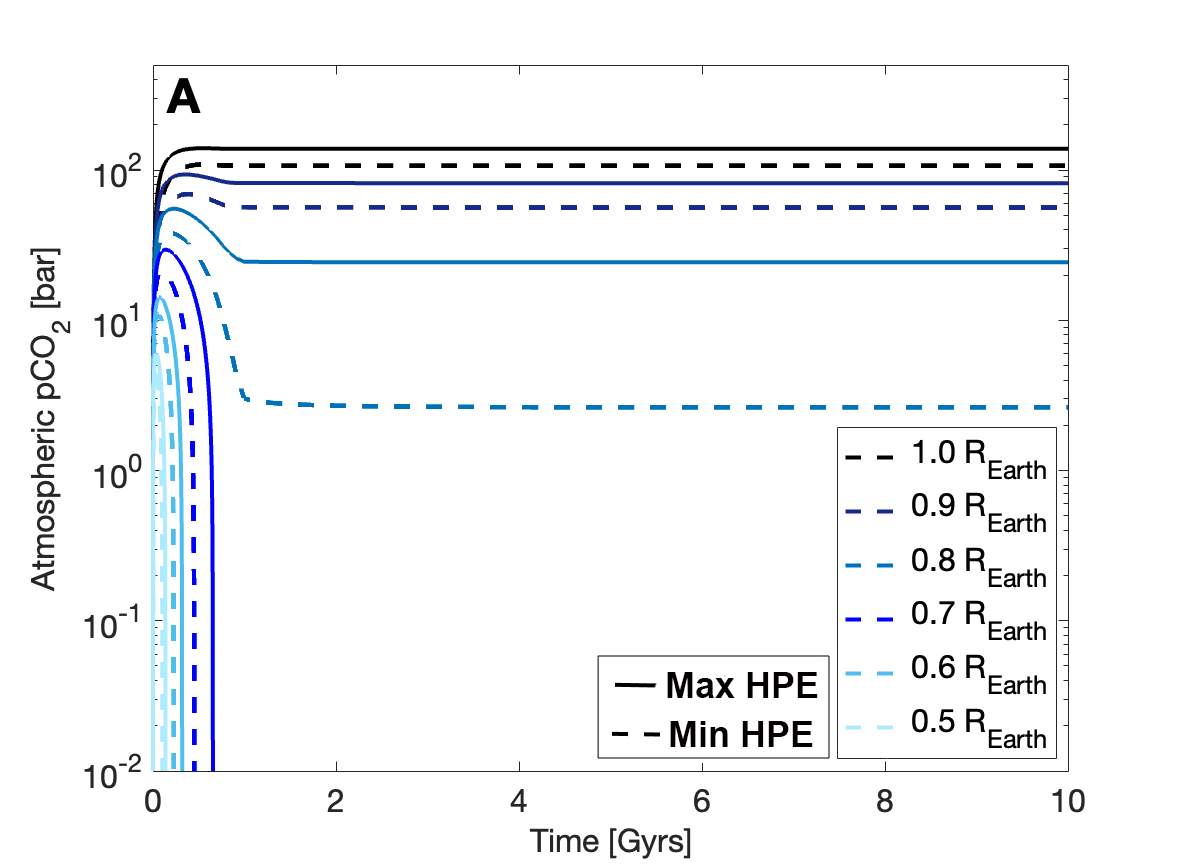}%
}
\subfloat{%
  \includegraphics[width=1\columnwidth, height=0.8\columnwidth]{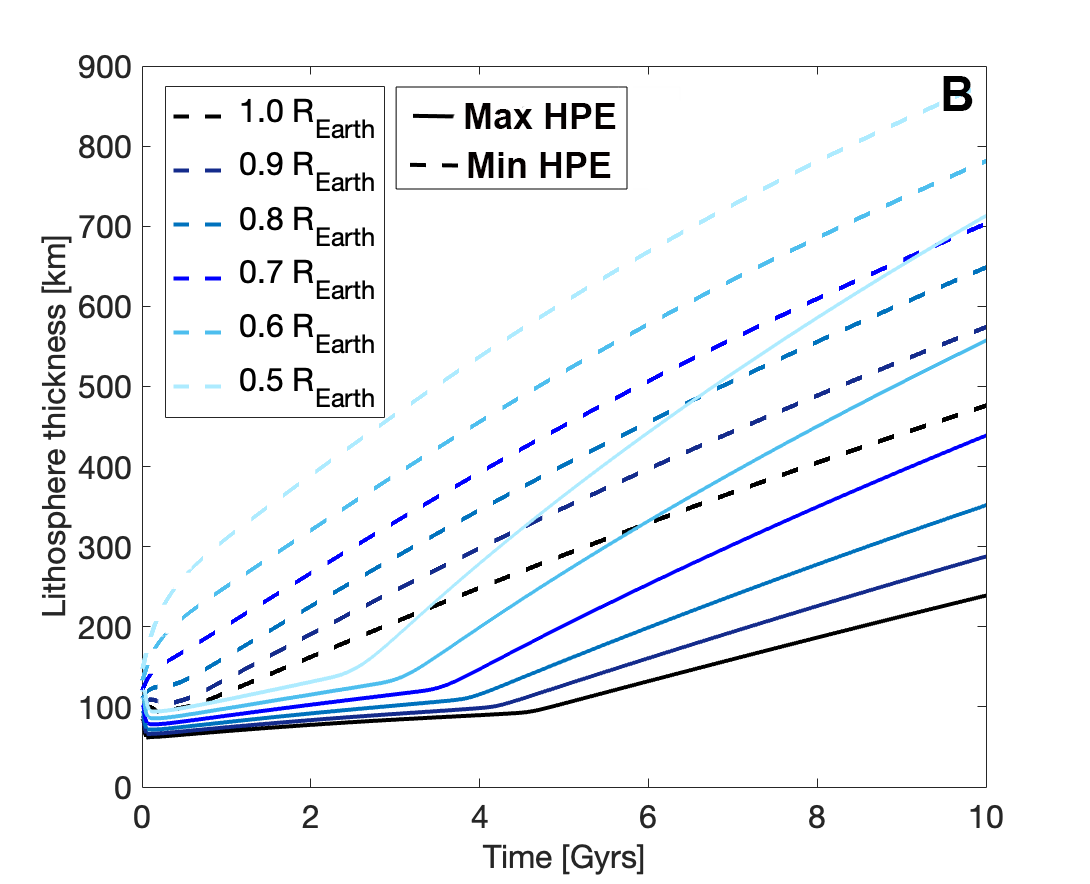}%
}\qquad
\subfloat{%
  \includegraphics[width=1\columnwidth, height=0.8\columnwidth]{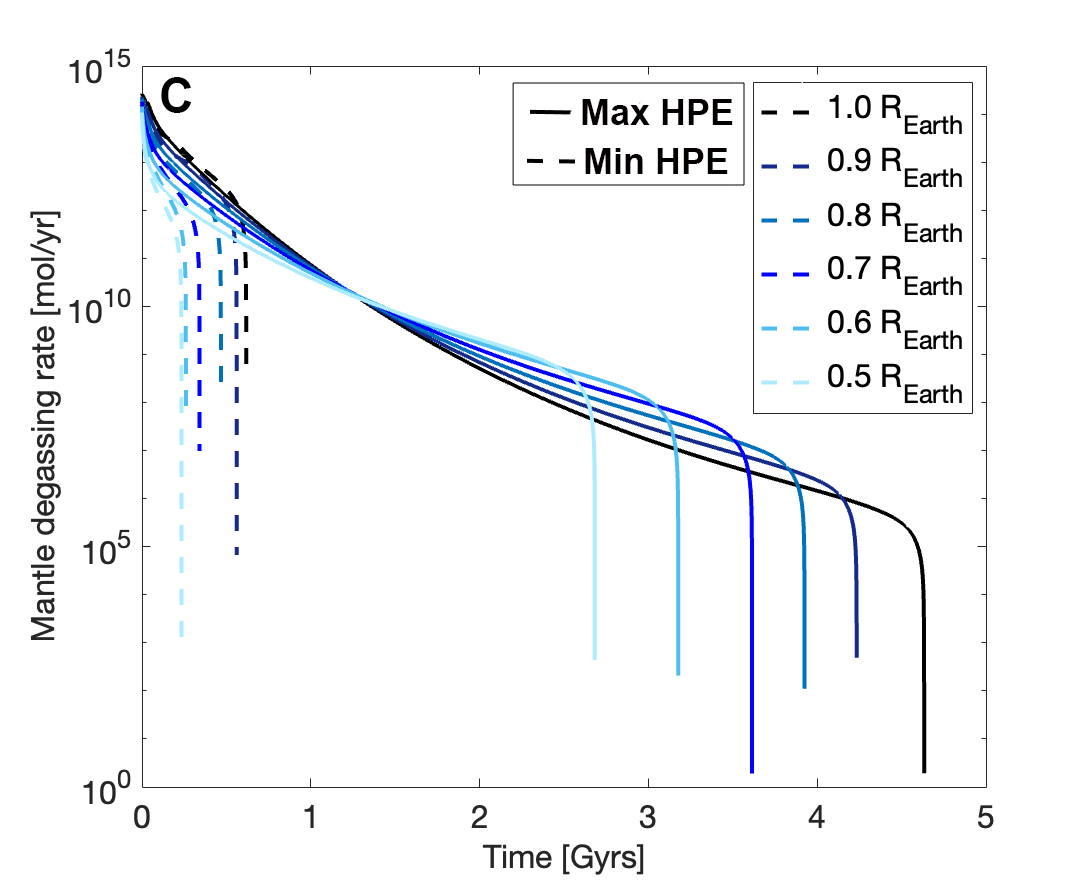}%
}
\subfloat{%
  \includegraphics[width=1\columnwidth, height=0.8\columnwidth]{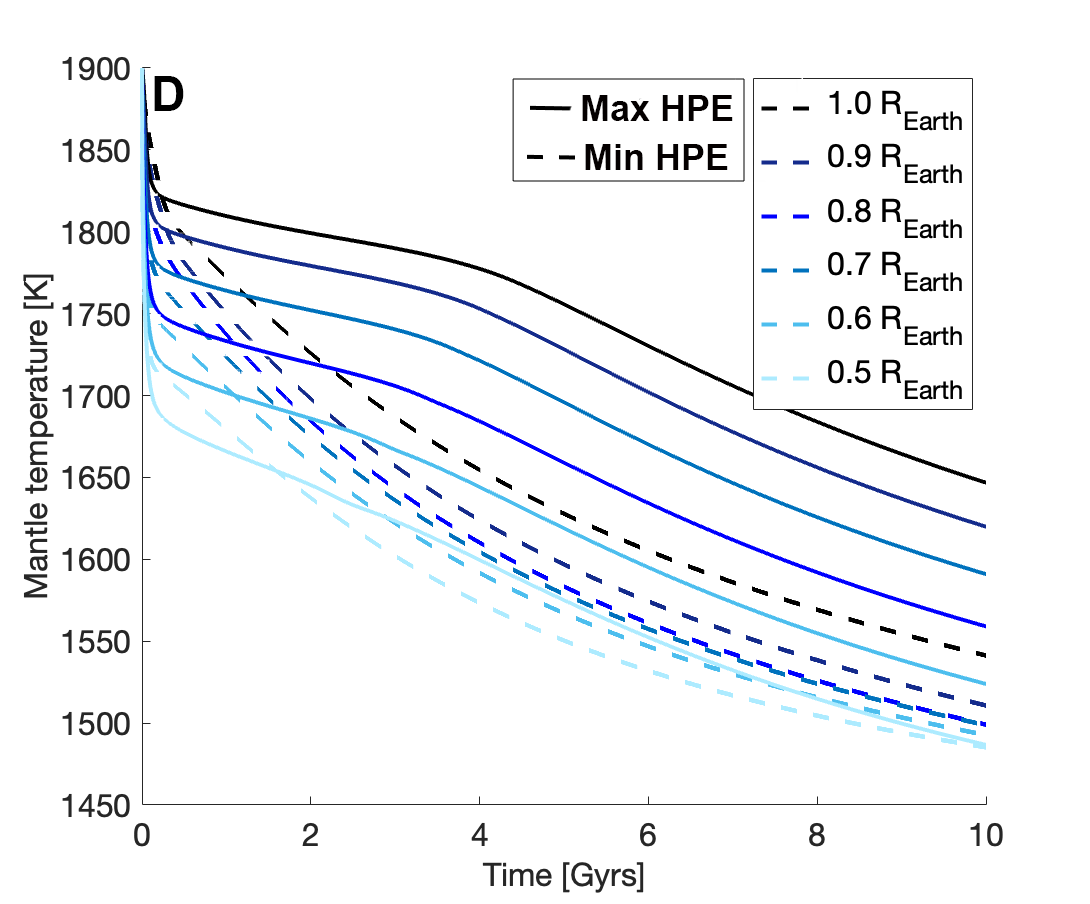}%
}
\caption{Variations in HPE Abundance. Other than the HPE abundances outlined in \ref{hpe}, the model uses default values set out in Section \ref{BradCode} for the remaining variables. Solid lines in each panel represent the maximum values for HPE, dashed lines represent minimum values of HPE.
The legend in each panel shows the color representative of each planet radius. Panel A shows the atmospheric CO$_2$ pressure. Both maximum and minimum values of HPE result in $\geq$ 0.8~R$_\oplus$ planets maintaining their atmosphere, though the low HPE model runs hold lower total atmosphere pressures than their high HPE counterparts. 
This is due to faster lithosphere growth (Panel B) and faster mantle cooling (Panel D) causing degassing to cease earlier (Panel C). 
\label{fig:HPE_vary}}
\end{figure*}

We explore the range of parameters outlined in Section \ref{BradCode} for the initial HPE abundance. Using the values from \citet{Unterborn2023} we use minimum values of (Th=Mg)$_{star}$ = 0.77 $\times$ Solar, (U/Mg)$_{star}$ = 0.45 $\times$ Solar, (K/Mg)$_{star}$ = 0.35 $\times$ Solar. For the low HPE planets the mantle degassing rate cuts off early (Figure \ref{fig:HPE_vary}, Panel C) due to the faster cooling of the mantle (Figure \ref{fig:HPE_vary}, Panel D) combined with a fast thickening of the lithosphere (Figure \ref{fig:HPE_vary}, Panel B). Planets $\geq0.8~R_\oplus$ maintain their atmosphere, however each size planet maintains a lower atmosphere pressure than the high HPE model runs, with the 1.0, 0.9 and 0.8~$R_\oplus$ planets holding onto 107, 56 and 3~bar atmospheres respectively. 
For model runs with the maximum value of HPE, we use HPE concentrations of (Th=Mg)$_{star}$ = 1.88 $\times$ Solar, (U/Mg)$_{star}$ = 1.92 $\times$ Solar, and (K/Mg)$_{star}$ = 3.63 $\times$ Solar. Planets $\geq0.8~R_\oplus$ again maintain their atmosphere (Figure \ref{fig:HPE_vary}, Panel A), with the 1.0, 0.9 and 0.8~$R_\oplus$ planets holding onto 139, 82 and 24~bar atmospheres respectively.
We find a larger amount of degassing occurs in the early stages compared to the default values of Figure \ref{fig:default_everything}. The mantle degassing for all planets lasts longer than the lower HPE runs, as well as the default runs (Figure \ref{fig:HPE_vary}, Panel C). This is due to a hotter mantle temperature for all planets at each age (Figure \ref{fig:HPE_vary}, Panel D) and a more slowly growing lithosphere (Figure \ref{fig:HPE_vary}, Panel B).

\setlength{\belowcaptionskip}{2pt}

\subsection{Variations in Initial Mantle Temperature}

\begin{figure*}
\centering
\subfloat{%
  \includegraphics[width=1\columnwidth, height=0.8\columnwidth]{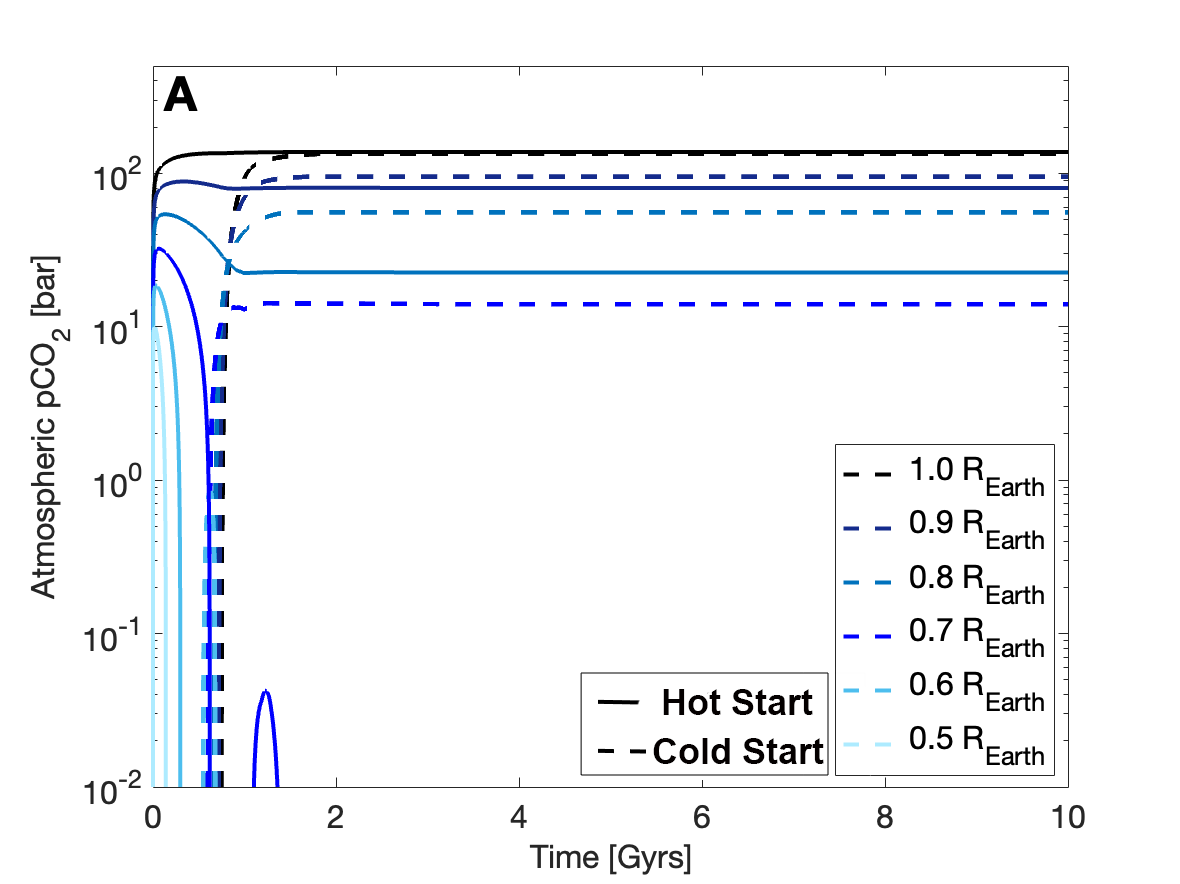}%
}
\subfloat{%
  \includegraphics[width=1\columnwidth, height=0.8\columnwidth]{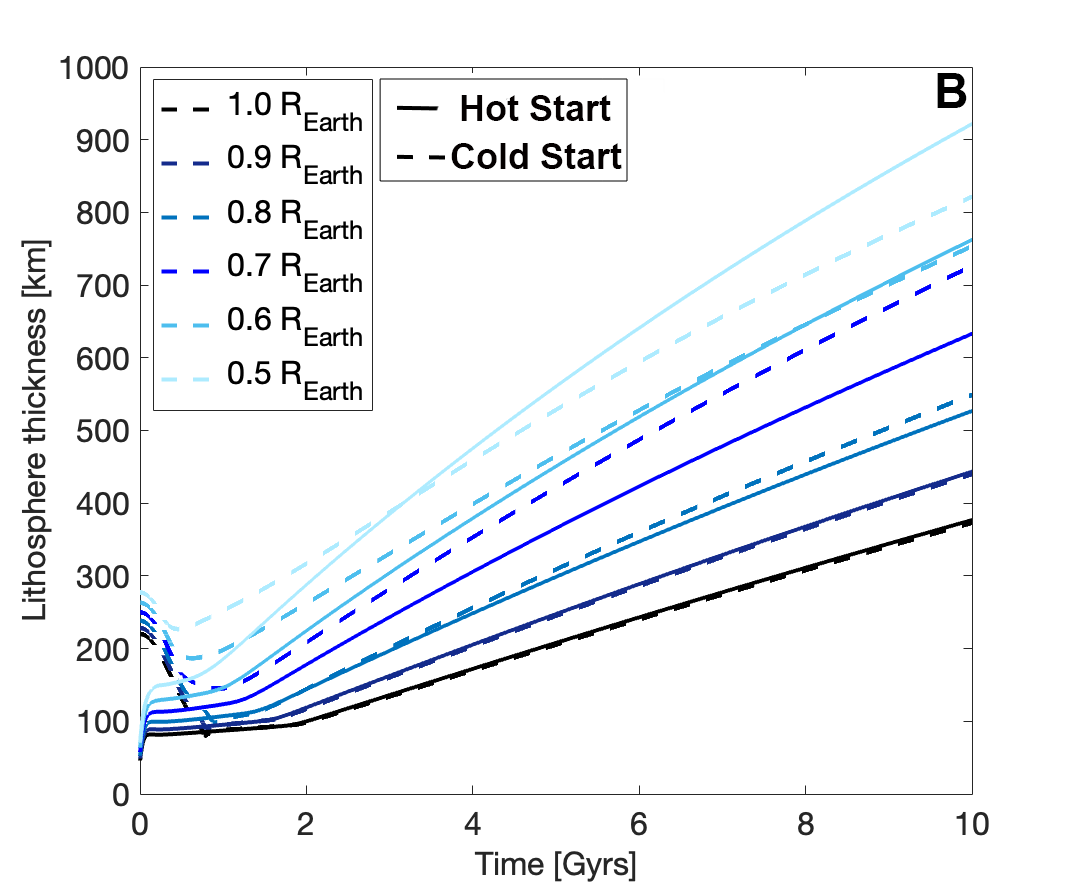}%
}\qquad
\subfloat{%
  \includegraphics[width=1\columnwidth, height=0.8\columnwidth]{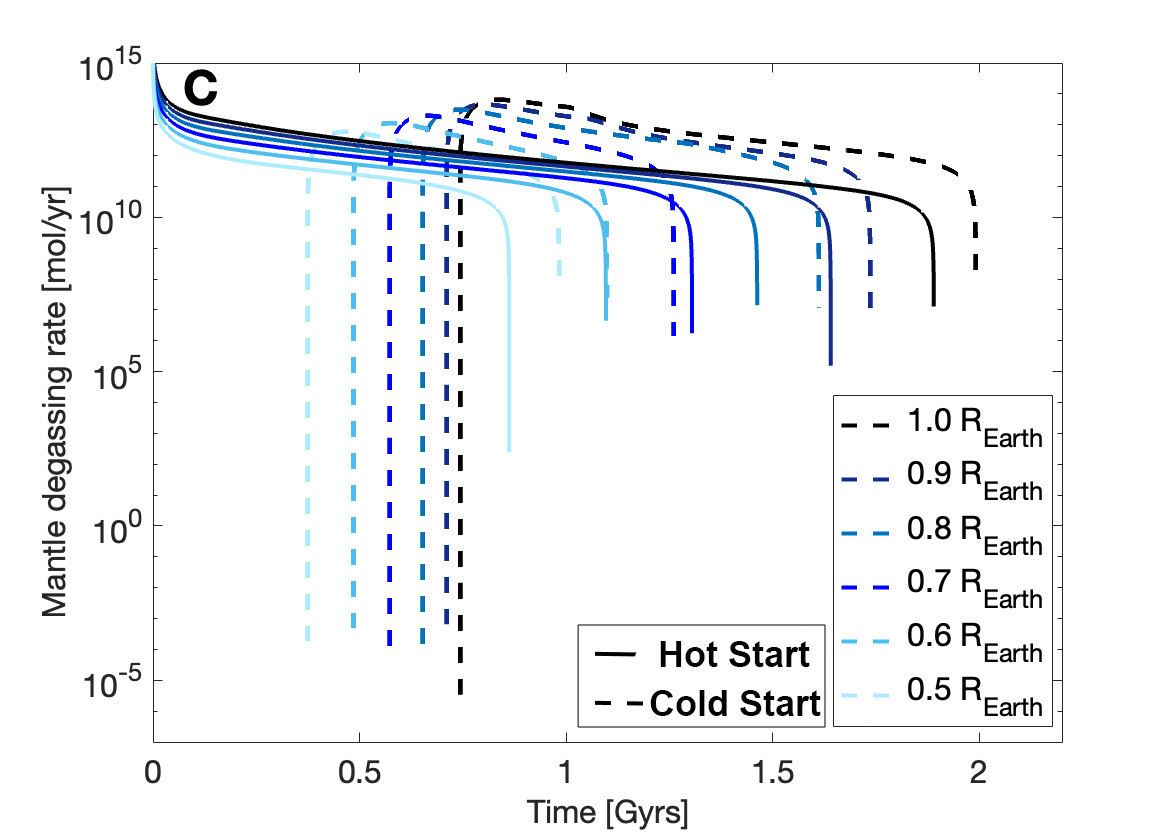}%
}
\subfloat{%
  \includegraphics[width=1\columnwidth, height=0.8\columnwidth]{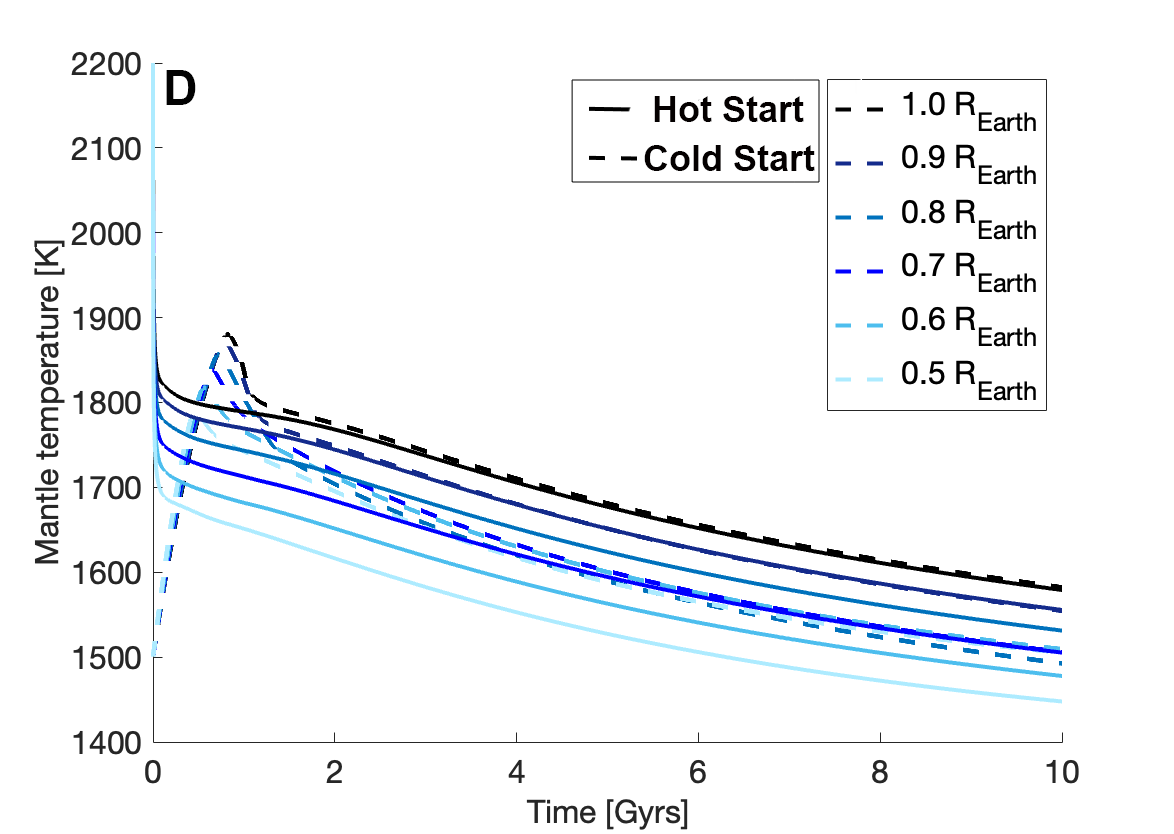}%
}
\caption{Variations in Initial Mantle Temperature. The maximum mantle temperature is 2200~K (hot start) shown with a solid line and the minimum mantle temperature is 1500~K (cold start) shown with a dashed line. The model uses default values set out in Section \ref{BradCode} for the remaining variables. In Panel A, all planets $\geq0.8~R_\oplus$ maintain their atmosphere for the hot starts, and all planets $\geq0.7~R_\oplus$ maintain their atmosphere for the cold starts. The cold start planets are unable to degas initially due to low mantle temperatures (Panel D) and 
an initially thick lithosphere (Panel B). The interior temperature increases over the first 0.5~Gyr (Panel D). During this time the lithosphere thickness reduces (Panel B) and melt is produced, allowing degassing to start (Panel C).  
\label{fig:man_temp_vary}}
\end{figure*}

We test how variation in initial mantle temperature will change the results, using a maximum mantle temperature of 2200~K and a minimum mantle temperature of 1500~K following the example of \citet{foley2019}. In Figure \ref{fig:man_temp_vary} the solid lines represent the model with the maximum starting mantle temperature, which we refer to as a ``hot-start", and dashed lines represent the minimum starting mantle temperature which we refer to as a ``cold-start". For the maximum initial mantle temperature runs, there is very little change from the default values. Planets of 1.0, 0.9 and 0.8~$R_\oplus$ hold onto 138, 80 and 23~bar atmospheres respectively. Planets $\leq0.7~R_\oplus$ lose theirs (Figure \ref{fig:man_temp_vary}, Panel A).  
Planets with a cold-start are better able to maintain an atmosphere, with planets as small as $0.7~R_\oplus$ maintaining a 14~bar atmosphere (Figure \ref{fig:man_temp_vary}, Panel A). This is due to the delay in outgassing that is caused by an early thick lithosphere (Figure \ref{fig:man_temp_vary}, Panel B) and an initially low mantle temperature (Figure \ref{fig:man_temp_vary}, Panel D). As the mantle temperature increases, melt is produced, lithosphere thins, and degassing begins (Figure \ref{fig:man_temp_vary}, Panel C). By this time the stellar XUV flux will have died down, allowing the carbon stores that were trapped in the mantle to build up in the atmosphere to greater pressures. However, while this low temperature start is important to include for completeness, it is not clear what physical process will produce initially cool mantle temperatures. If planet formation leads to significant heating, then planets might mostly have warm starts, and so this cold start scenario may not be realistic.

\setlength{\belowcaptionskip}{2pt}

\subsection{Variations in Planet Density}
\label{density}

\begin{figure}
\begin{center}
\includegraphics[width=0.5\textwidth]{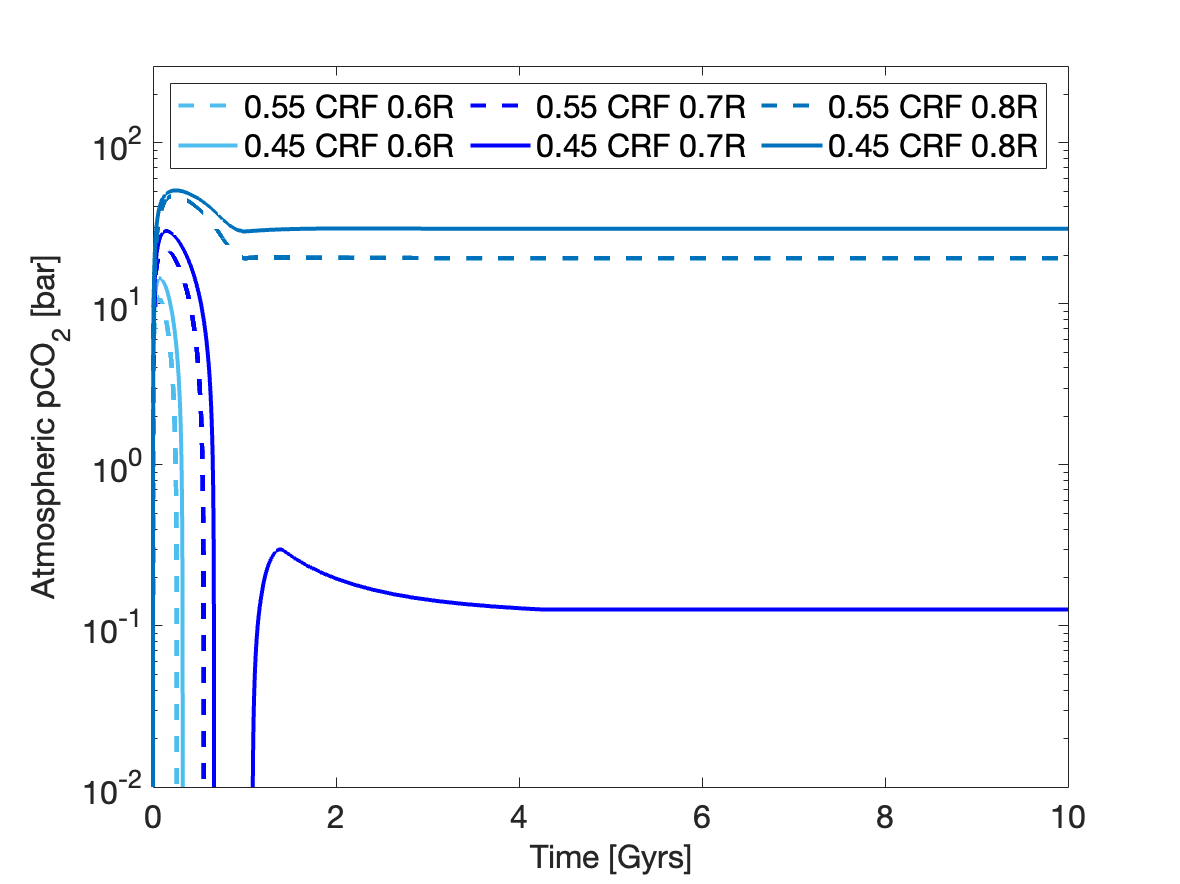}
  \end{center}
\caption{Variations in CRF set in ExoPlex within the bounds of code capabilities produced different results for the atmospheric CO$_2$ pressure retained. A change of $\sim$10~bar is seen in the 0.8$~R_\oplus$ planets, and a regained atmosphere of 0.1~bar for the low CRF 0.7$~R_\oplus$ planet, whereas the high CRF 0.7$~R_\oplus$ planet never regained its atmosphere. Both 0.6$~R_\oplus$ planets lost their atmospheres.  
\label{fig:Exoplex}}
\end{figure}

\begin{figure*}
\centering
\subfloat{%
  \includegraphics[width=1\columnwidth, height=0.8\columnwidth]{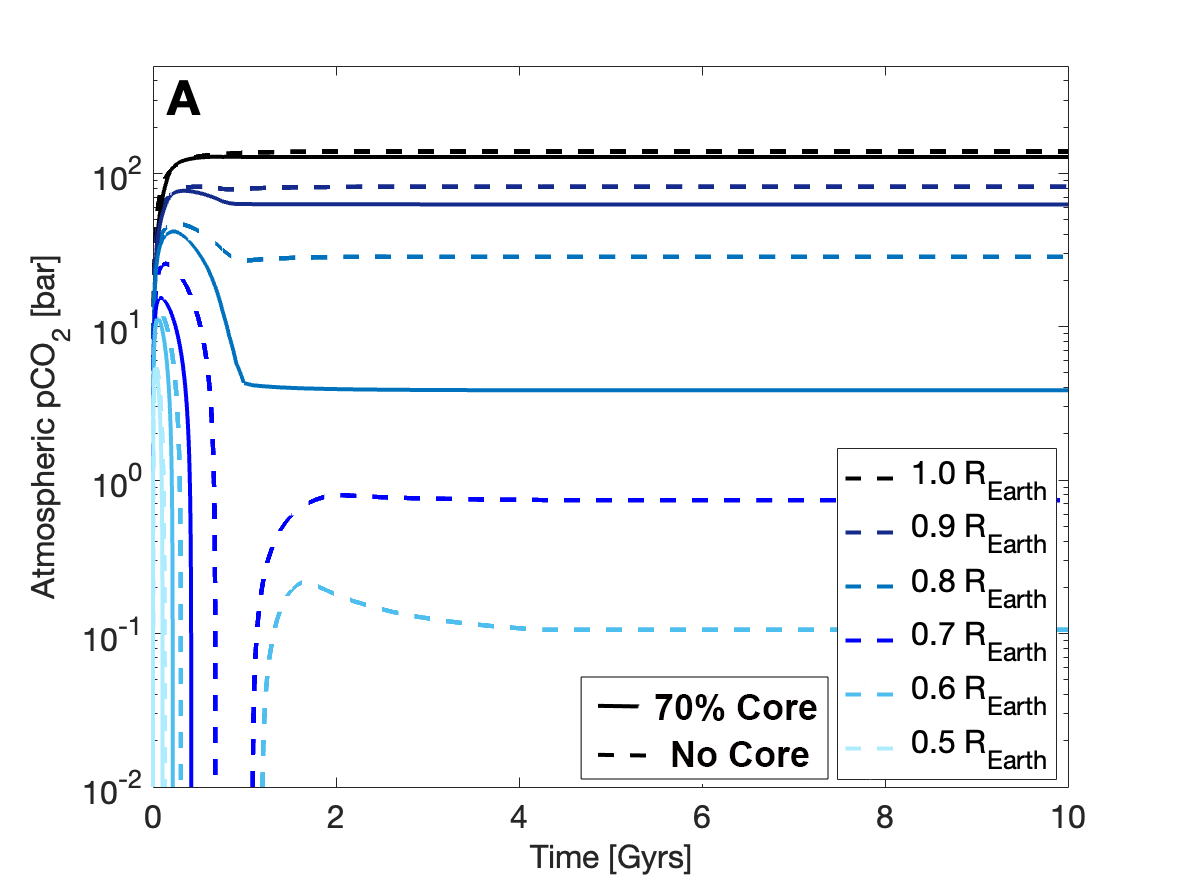}%
}
\subfloat{%
  \includegraphics[width=1\columnwidth, height=0.8\columnwidth]{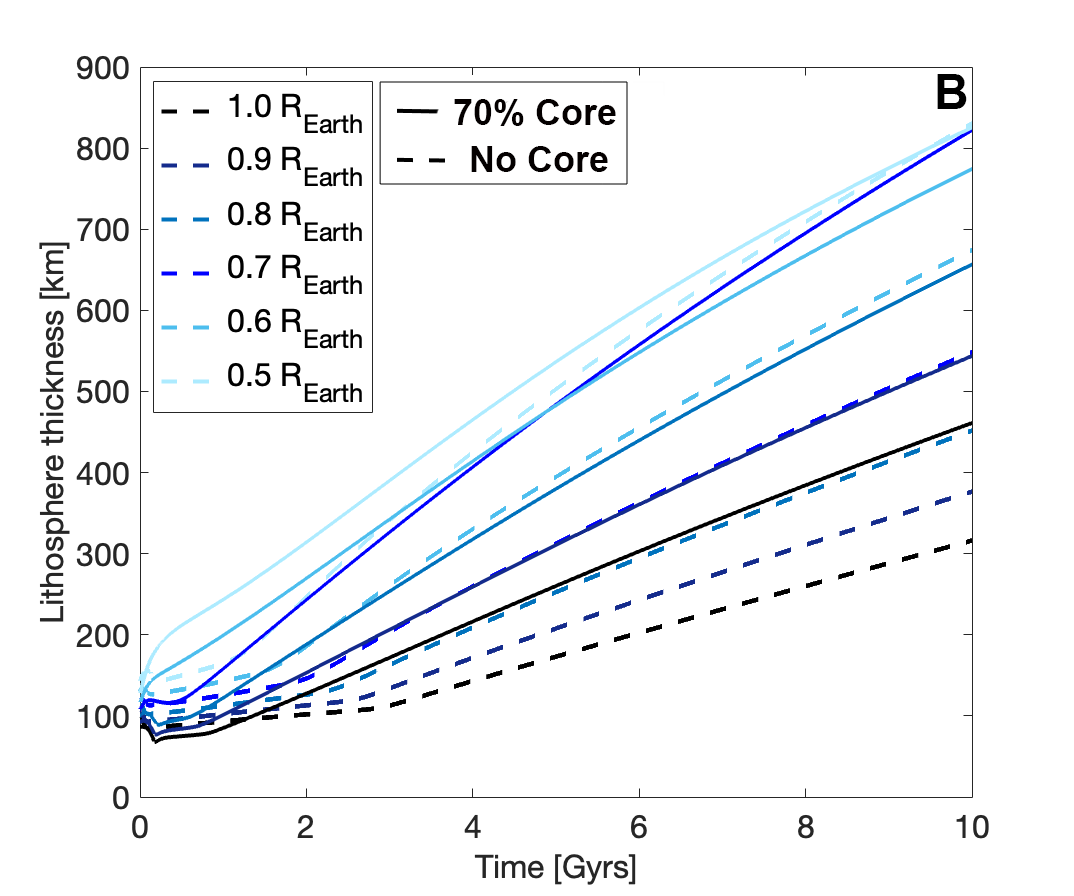}%
}\qquad
\subfloat{%
  \includegraphics[width=1\columnwidth, height=0.8\columnwidth]{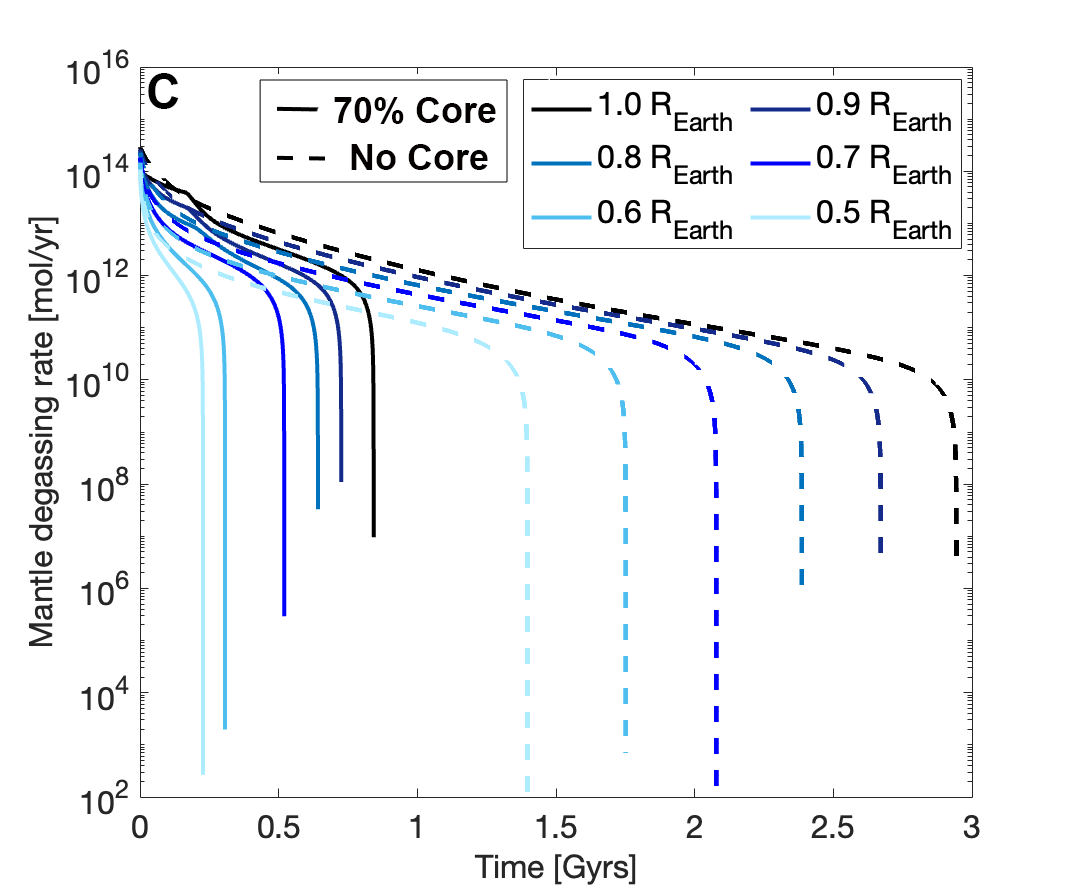}%
}
\subfloat{%
  \includegraphics[width=1\columnwidth, height=0.8\columnwidth]{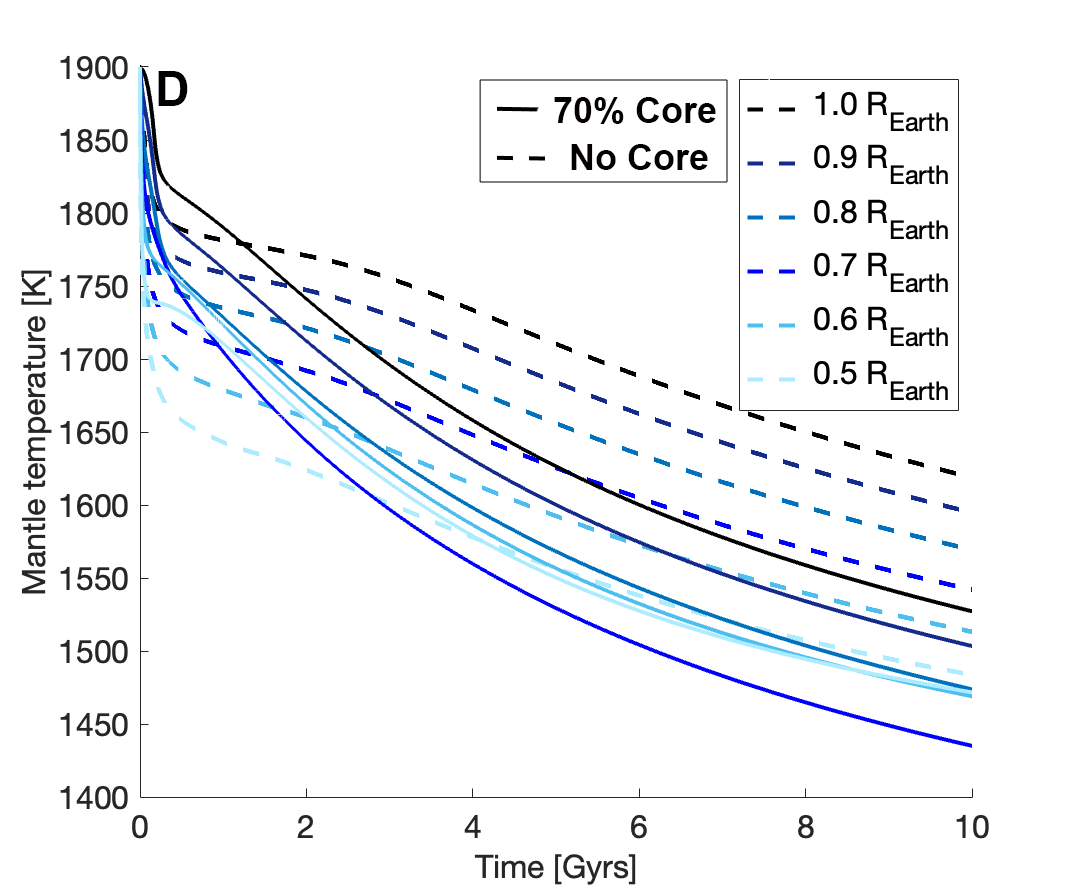}%
}
\caption{Variations in Planet Density. The panels in this figure show the results of when CRF, and thus planet density, is varied. The model uses default values set out in Section \ref{BradCode} for the remaining variables. We tested planets with no core (dashed lines) and planets with a 0.70 CRF (solid lines). For both the planets with a large core and no core, planets $\geq~$0.8~$R_\oplus$ maintained their atmosphere. For the planets with no core, planets 0.7 and 0.6~$R_\oplus$ regained their atmospheres once XUV flux reduced to a level that allowed CO$_2$ pressure to rebuild.  
\label{fig:density_vary}}
\end{figure*}

We tested the effect that a variation of planet density has on the model results by changing the CRF from the default value of 0.55. We first tested how changes within the bounds of Exoplex capabilities (0.55-0.44 CRF, determined by adding up to 20\% by mass of FeO into the mantle) produced slightly different results, with a variation of 10~bar for the 0.8$~R_\oplus$ planet, and both 0.6$~R_\oplus$ planets losing their atmospheres (Figure \ref{fig:Exoplex}). However, a change occurs for the 0.7$~R_\oplus$ planet, with a regained atmosphere of 0.1~bar for the low CRF planet, whereas the high CRF 0.7$~R_\oplus$ planet never regained any atmosphere.  
We modeled planets with larger changes in CRF, ranging from planets with no core representing our minimum density planets, to planets with a CRF of 70\% representing our maximum density planets. To calculate surface gravity for the ``No core" planets we took the average mantle and core densities for each planet size from the min ExoPlex CRF of 0.45. For the ``70\% core" planets we took the average mantle and core densities for each planet size from the maximum ExoPlex CRF of 0.55. We did not increase the CRF above 0.70 as increasing the core size had the effect of reducing the mantle volume and a corresponding reduction in volatile inventory and HPE, as carbon and HPE are scaled with the mantle mass. This caused a reduction in atmosphere retention for the planets with large cores. Thus Figure~\ref{fig:density_vary}, Panel A shows that for both the planets with a large core and those with no core, $\geq~$0.8~$R_\oplus$ planets maintained their atmosphere (Figure~\ref{fig:density_vary}, Panel~A). For the planets with no core, however, greater atmospheric pressure was maintained on each planet, and the 0.7 and 0.6~$R_\oplus$ planets were able to regain atmospheres of 0.7 and 0.1~bar each once solar wind had reduced so that CO$_2$ pressure was able to rebuild. No core planets of 1.0, 0.9 and 0.8~$R_\oplus$ held onto 139, 82 and 28~bar atmospheres respectively, whereas planets with 0.7~CRF held onto 128, 63 and 4~bar atmospheres for the 1.0, 0.9 and 0.8~$R_\oplus$ planets respectively. Provided the HPE and volatiles scale with mantle mass, a no core planet will provide greater opportunity for a small planet to maintain an atmosphere, and provide sufficient volatile and HPE inventory that a planets lost atmosphere may be recovered after the XUV flux has reduced. Note, we also tested the ``70\% core" planets with 1.5$\times$ mantle density and corresponding larger surface gravity to account for the approximation made by using the 0.55 ExoPlex density results and found that again the planets $\geq$0.8~$R_\oplus$ maintained their atmosphere.  

Even though the larger core provides a larger surface gravity which helps prevent atmosphere loss, the smaller mantle volume and corresponding reduction in volatile materials and HPE has a greater effect. Reduced HPE allows mantle cooling to occur more rapidly (Figure~\ref{fig:density_vary}, Panel~D), the lithosphere to grow faster (Figure~\ref{fig:density_vary}, Panel~B), and degassing to shutoff sooner (Figure~\ref{fig:density_vary}, Panel~C), all contributing to the reduced total CO$_2$ pressure for each planet size (Figure~\ref{fig:density_vary}, Panel~A).

As the model does not include core cooling, variations to the model results are expected. Mantle heat may remain steady, as results from \citet{DRISCOLL2014} show a similar thermal evolution and contribution to the mantle heat from both core heat flow across the core-mantle-boundary (CMB) and internal heat flow from mantle HPE. Carbon content in the mantle, however, will still be reduced with a larger CRF, and as seen in Section \ref{carbon}, initial carbon content is one of the greater influences on a planet's ability to hold onto an atmosphere.

\setlength{\belowcaptionskip}{2pt}

\subsection{The Habitable Zone boundaries}
\label{HZ}

\begin{figure}
\begin{center}
\includegraphics[width=0.5\textwidth]{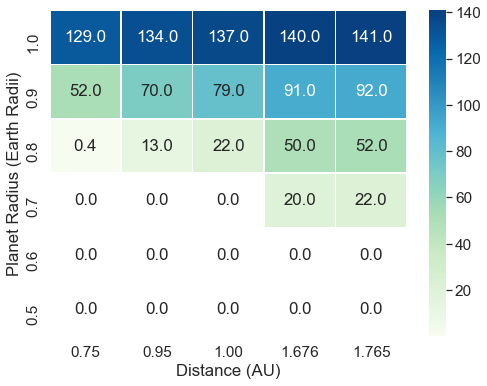}
  \end{center}
\caption{A heat map of atmosphere retention for planets from 1.0--0.5~$R_\oplus$ at distances that represent each of the HZ boundaries around a Sun-like star, along with the default value of 1 AU. Atmospheric pressure is depicted by the colorbar along with the values in each grid, and measured in bar. The value shown is the final atmospheric pressure when both atmosphere loss and degassing has shutdown on the planet.
\label{fig:HZ_boundaries}}
\end{figure}

Using the calculations outlined in Section~\ref{verify} we determined the flux received at the inner and outer edges of the HZ for a sun like star in accordance with the HZ boundaries defined by \citet{kopparapu2013a,kopparapu2014}. We used both the conservative habitable zone (CHZ) boundaries of Runaway Greenhouse limit at 0.95 AU and the Maximum Greenhouse limit at 1.676 AU and the optimistic HZ boundaries (OHZ) of the Recent Venus limit at 0.75 AU and the Early Mars limit at 1.765 AU. The corresponding relative flux values were 1.107--0.356~F$_\odot$ for the CHZ and 1.776 - 0.321~F$_\odot$ for the OHZ. 
At the outer edge of both the OHZ and CHZ, or the early Mars limit and maximum greenhouse limit, all planets $\geq$0.7~$R_\oplus$ are able to maintain their atmosphere (Figure \ref{fig:HZ_boundaries}). This is due to both the lower amount of flux received over time, as well as the flux received dropping below the limit at which CO$_2$ is readily stripped earlier. At the inner edge of the OHZ and CHZ, or the recent Venus limit and runaway greenhouse limits, planets $\geq$0.8~$R_\oplus$ are able to maintain their atmosphere, though only a tenuous 0.4~bar atmosphere remains on the 0.8~$R_\oplus$ planet at the recent Venus limit. Note, \citet{kopparapu2014} showed that the habitable zone boundaries have mass dependence. A low mass planet has a HZ inner edge boundary at a greater distance from the star, while the outer edge remains fairly stable. For a 0.5~$R_\oplus$ planet, the CHZ inner boundary is at 1.005~AU. This is close to our default distance of 1~AU and we see no significant change to the atmosphere retention of the planets with this minor difference. At this mass adjusted Runaway Greenhouse limit, planets $\geq$0.8~$R_\oplus$ maintain their atmosphere.

\subsection{Exobase Temperatures}
\label{Exobase}

We tested exobase temperatures of 400~K, 800~K and 2500~K with otherwise default conditions. Although a thick CO$_2$ atmosphere will provide strong cooling to the upper atmosphere, we tested this range to determine the effect of higher exobase temperatures on atmosphere retention. Higher exobase temperatures led to a decrease in total atmospheric pressure. For the 800~K exobase temperature model runs, planets 0.8~$R_\oplus$ and above still maintained their atmospheres. For the 2500~K exobase temperature model runs, only planets 0.9~$R_\oplus$ and above maintained their atmospheres while 0.8~$R_\oplus$ and below lost theirs.

%%%%%%%%%%%%%%%%%%%%%%%%%%%%%%%%%%%%%%%%%%%%%%%%%%%%%%%%%%%%%%%%%%%%

\section{Discussion}
\label{Disc}

Our modeling results suggest that for low-mass planets, atmospheric retention is optimized with specific initial conditions and physical parameters. A cold initial mantle temperature proves advantageous, as it delays carbon outgassing until stellar activity has diminished, thereby reducing the risk of atmospheric loss through stellar XUV flux erosion. Higher initial mantle carbon budgets are beneficial for atmospheric replenishment through volcanic outgassing. Higher concentrations of HPEs allow for longer degassing, with slightly higher atmosphere pressures than the default model. The model further indicates that a lower core mass fraction is favorable, provided HPE and volatile material abundance scales with the size of the mantle. 
Planets positioned away from the inner edge of the HZ benefit from reduced stellar flux, which minimizes atmospheric escape. While the HZ boundary simulations indicate that planets closer to the outer edge of the HZ are able to hold onto their atmospheres more easily, at these larger orbital distances, a planet needs a more substantial greenhouse effect to achieve temperatures suitable for liquid water. It was shown in \citet{Schwieterman2019} that the levels of CO$_2$ required to maintain habitable temperatures on the surface of planets in the outer HZ may prevent complex life from surviving there.

While many of the stagnant-lid model runs produce planets with thick CO$_2$ atmospheres, our model does not include CO$_2$ draw down through weathering as we focused on maximizing retention of CO$_2$ in the atmosphere. Provided water was available, \citet{FoleySmye18} found that 0.01--1$\times$Earth's carbon inventory could be maintained at habitable conditions through weathering. Providing weathering does not become limited by the lack of fresh basalt or water, stagnant lid planets can prevent excessive CO$_2$ buildup.

The stellar XUV history is a major source of uncertainty in atmospheric escape calculations, as estimates for the young Sun span more than an order of magnitude depending on its rotational evolution and the normalization of empirical reconstructions. The widely used relation from Ribas et al. (2005) is based on solar analogs and provides a robust description of the decay in XUV flux, but in its original form overestimates the present-day solar XUV flux. To ensure consistency with the modern Sun, we therefore apply a uniform renormalization, reducing the Ribas et al. flux at all ages by a factor of 4.5 while preserving its time dependence. In addition, rather than adopting the upper-envelope XUV histories implied by rapidly rotating young stars, we impose a maximum flux of 10× the present-day solar value. These assumptions represent a conservative XUV evolution scenario that avoids overestimating atmospheric loss rates while remaining consistent with observational constraints and the range of plausible solar histories. If the true early solar XUV flux were higher than assumed here, our model would predict enhanced atmospheric escape, shifting the threshold for long-term atmosphere retention toward larger planetary sizes.

Beyond the thermal atmosphere escape mechanisms used in this model, non-thermal processes can significantly increase atmospheric loss rates beyond what these 1D thermal escape calculations predict \citep{jakosky2018,dong2018b,Lichtenegger2010,LICHTENEGGER2022}. The interaction between a planet's atmosphere and its host star's magnetic field can drive ion pickup processes, stripping away atmospheric particles that become ionized. Sputtering, where energetic particles impact the upper atmosphere and eject neutral particles, provides another loss pathway. Additionally, coronal mass ejections, which are not included in the model, can strip away atmospheric particles through their intense particle and radiation fluxes. For smaller planets with lower surface gravity, these non-thermal processes can dominate the atmospheric loss budget, and will persist after the early, highly active stage of the star has reduced. Consequently, our model results represent an optimistic scenario for atmospheric retention. In reality, smaller planets would likely experience more severe atmospheric loss due to these additional non-thermal processes. 
Also, as mentioned in Section~\ref{Atmesc}, CO$_2$ is a heavy molecule, which combined with its radiative cooling effects can make it a difficult molecule to lose. We focus on a pure CO$_2$ atmosphere to represent a best-case scenario for atmosphere retention.

The timing of outgassing could impact atmospheric retention in ways not fully captured by our continuous outgassing model. Episodic outgassing could substantially alter atmospheric evolution, particularly if major outgassing episodes occur after the host star's activity has diminished. Late-stage outgassing events would benefit from reduced stellar XUV flux, allowing a greater fraction of the outgassed CO$_2$ to be retained in the atmosphere. Additionally, episodic outgassing could lead to periods of rapid atmospheric pressure buildup, potentially creating temporary conditions favorable for liquid water stability even on smaller planets. This suggests that planets with delayed or episodic outgassing patterns might retain larger atmospheres longer than predicted by our continuous outgassing models.
Late-stage impacts could also provide a mechanism for a regained atmosphere and water delivery, particularly if they occur after the most intense period of stellar activity has subsided. 

In Figure~\ref{fig:density_vary} the 0.7 and 0.6~$R_\oplus$ planets with no core lose their atmospheres but are later able to regain 0.7 and 0.1~bar atmospheres respectively. Provided planets have sufficient volatiles and HPE to produce melt and degas, planets may be able to recover their atmospheres once XUV flux levels reduce enough to allow atmosphere CO$_2$ to build up. If water reservoirs can survive through the low pressure phase as subsurface ice, buried aquifers, or partially retained surface oceans, then liquid water may also re-emerge following atmospheric recovery. However, the survivability of water during a transient period of atmospheric loss remains uncertain. While studies have looked into the loss of oceans of water in low atmospheric pressure through evaporation and escape \citep{wordsworth2013b}, how large an initial water inventory is required to avoid complete ocean loss during a temporary period of low pressure remains poorly constrained. It is unclear whether a $\leq$0.7~$R_\oplus$ planet would be able to retain the necessary volatile inventory.
Future studies should investigate the minimum water reservoir needed to survive atmospheric collapse and subsequent recovery for these small planets.

Our model neglects the effect of an earlier magma ocean period on the intial volatile budget of the planet. In the field of planet formation, it is often assumed that all rocky planets go through at least one episode of magma ocean during their formation \citep{elkinstanton2012} but it remains unclear whether all small rocky objects go through periods of complete melt. For instance, Mars is small enough that early heat escape is efficient and only rapid accretion times with significant atmospheric greenhouse warming would produce fully molten mantles \citep{saito2018}. Magma ocean cooling times for habitable zone rocky planets around Sun-like stars are on the order of 1-3 Myr, much shorter than magma ocean lifetimes around smaller M-dwarf stars, although magma oceans for Venus-like insolation levels can persist up to 100 Myr \citep{lebrun2013,hamano2013,salvador2017}. Melt-trapping during magma ocean crystallization can lead to the sequestration of up to 80\% of total water under Earth-like conditions and up to 80\% of total carbon in the mantle under reducing, Mars-like conditions \citep{hiermajumder2017,sim2024}, although carbon trapping can be  considerably lower (1-10\%) for oxidizing conditions. Post-magma ocean weathering processes are expected to be rapid \citep{zahnle2010}, especially when an early warm ocean forms out of an initial steam atmosphere, on the order of 20-100 Myr for total drawdown of 90 bars of CO$_2$. Although an early magma ocean period produces some uncertainty on the initial distribution of carbon between planetary reservoirs, the relatively short timescale of magma oceans around Sun-like stars indicates that significant atmospheric escape during this time period is not likely to play a major role in the total carbon budget, with the possible exception of the innermost planets that we consider here.   

Our model excludes the effects of planetary magnetic fields, as their role in atmospheric retention remains a subject of debate. While traditional views suggested that magnetic fields act as protective shields against atmospheric loss \citep{Brain2013,Driscoll2015}, recent studies have presented compelling evidence that magnetic fields might actually enhance atmospheric escape \citep{Sakai2018} or may vary with the size of the field \citep{hinton2024}. Given this ongoing uncertainty in the scientific community and the computational complexity of accurately modeling magnetic field effects, we chose to omit magnetic fields from our current analysis, acknowledging that their inclusion could either enhance or diminish the atmospheric retention rates presented in our results.

%%%%%%%%%%%%%%%%%%%%%%%%%%%%%%%%%%%%%%%%%%%%%%%%%%%%%%%%%%%%%%%%%%%%

\section{Conclusions}
\label{conclusions}

Our research demonstrates that atmospheric retention, a crucial factor in maintaining habitable conditions, varies significantly with the size of the planet.
Using default Earth-like parameter values, stagnant lid planets $\geq$0.8~$R_\oplus$ can maintain their atmospheres throughout their evolution. 

Among the various parameters investigated, the initial carbon inventory emerges as the most influential factor in determining atmospheric retention. When carbon is increased to the maximum value estimated for Earth, planets $\geq$0.8~$R_\oplus$ are able to maintain their atmospheres, while with an order of magnitude more carbon, all but the smallest 0.5~$R_\oplus$ planet maintained their atmosphere. Planets with no core also demonstrate enhanced atmospheric retention capability, with planets $\geq$0.8~$R_\oplus$ maintaining their atmospheres, and the 0.7 and 0.6~$R_\oplus$ planets regaining their lost atmospheres once stellar XUV flux diminishes. The magnitude of these atmospheres also shows a clear relationship with planet size, with a 0.7~bar atmosphere regained for the 0.7~$R_\oplus$ planet and a 0.1 bar regained atmosphere for the 0.6~$R_\oplus$ planet.
A cooler initial mantle temperature of 1500K favors atmospheric retention by delaying outgassing until after peak stellar activity, allowing planets $\geq$0.7~$R_\oplus$ to maintain their atmospheres. However, as planet formation leads to significant heating, this low temperature start may not be realistic. Position within the HZ also impacts retention, with planets near the outer edge better maintaining atmospheres due to the reduced stellar flux . However, while planets orbiting farther from their stars might seem safer from intense stellar radiation and activity, maintaining warm enough surface temperatures at greater distances requires atmospheres that could become toxic \citep{Schwieterman2019}. 

Our results indicate that the most favorable conditions for atmospheric retention on sub-Earth-sized planets involve a combination of high initial mantle carbon content, low CRF, cool initial mantle temperature, high HPE concentrations, and an orbital position away from the inner edge of the HZ, but not far enough that the atmosphere would be toxic. A combination of some of these factors would create optimal conditions for small planets to maintain substantial atmospheres over geological timescales.
These findings have important implications for identifying potentially habitable exoplanets. 

While smaller planets face greater challenges in maintaining atmospheres, our model suggests they can redevelop atmospheres under the right conditions. This is particularly relevant for future observations of sub-Earth-sized planets, indicating that even those that initially lose their atmospheres should not be immediately discounted as potentially habitable worlds.
The model emphasizes the importance of considering a planet's full evolutionary history rather than just its current state. The timing and nature of outgassing, early stellar activity, and initial conditions all play crucial roles in determining a planet's long-term atmospheric stability. This understanding will be vital for interpreting future atmospheric observations of small rocky exoplanets and assessing their potential habitability.

During initial testing we also found the model predicts Earth as a stagnant lid planet would have gained a 126 bar CO$_2$ atmosphere. This indicates that an Earth sized stagnant lid planet at Earth’s current position within the HZ of a sun-like star would be uninhabitable and dominated by a dense CO$_2$ envelope.
In fact, many of the stagnant lid planet model runs resulted in planets with thick CO$_2$ atmospheres which would be potentially incompatible with life. Planet's with plate tectonics that enable recycling of CO$_2$ back into the mantle may be necessary to maintain habitable levels of atmospheric CO$_2$.
Future research directions will explore the effects of plate tectonics on maintaining habitable conditions, as well as additional factors not included in our current model such as M-dwarf stellar environments, tidal locking, and tidal heating. These factors could further modify the conditions under which small planets can maintain substantial atmospheres and potentially host life. This expanded understanding will be crucial as we continue to search for and characterize potentially habitable worlds beyond our Solar System.

%%%%%%%%%%%%%%%%%%%%%%%%%%%%%%%%%%%%%%%%%%%%%%%%%%%%%%%%%%%%%%%%%%%%

\section*{Acknowledgements}

M.L.H. would like to acknowledge NASA support via the FINESST Planetary Science Division, NASA award number 80NSSC21K1536. M.L.H. would also like to  thank the Stanford Science Fellowship. We would like to thank Cayman Unterborn for his advice regarding the use of ExoPlex, and the reviewers whose comments and questions have strengthened the quality and clarity of this work.

%%%%%%%%%%%%%%%%%%%%%%%%%%%%%%%%%%%%%%%%%%%%%%%%%%%%%%%%%%%%%%%%%%%%

%%%%%%%%%%%%%%%%%%%%%%%%%%%%%%%%%%%%%%%%%%%%%%%%%%%%%%%%%%%%%%%%%%%%

\end{document}